\documentclass[aps,pra,superscriptaddress,twocolumn,notitlepage,10pt]{revtex4-1}

\usepackage{graphicx,graphics,times,color}
\usepackage{bm}
\usepackage{longtable}
\usepackage{color}
\usepackage{lipsum}
\usepackage{epstopdf}
\usepackage{amsmath}
\usepackage{algcompatible}

\mathchardef\minus="002D

\newcommand{\ket}[1]{\left|{#1}\right\rangle}
\newcommand{\bra}[1]{\langle{#1}|}

\newcommand\V[1]{\boldsymbol{#1}}
\DeclareMathOperator*{\Tr}{Tr}

\begin{document}

\title{Quantum gate learning in qubit networks: Toffoli gate without 
time dependent control} 

\author{Leonardo Banchi}
\affiliation{Department of Physics and Astronomy, University College London, Gower Street, WC1E 6BT London, United Kingdom}

\author{Nicola Pancotti}
\affiliation{Max-Planck-Institut f\"ur Quantenoptik, Hans-Kopfermann-Stra{\ss}e 1,
85748 Garching, Germany}
\affiliation{Dipartimento di Fisica, Sapienza Universit\`a di Roma, Piazzale Aldo
Moro, 5, I-00185 Roma, Italy}

\author{Sougato Bose}
\affiliation{Department of Physics and Astronomy, University College London, Gower Street, WC1E 6BT London, United Kingdom}

\date{\today}

\begin{abstract}
We put forward a strategy to encode a quantum operation into the 
unmodulated dynamics of a quantum network without the need of external 
control pulses, measurements or active feedback. 
Our optimization scheme, inspired by supervised machine learning, 
consists in engineering the pairwise couplings between the network qubits 
so that the target quantum operation is encoded in the natural 
reduced dynamics of a network section. 
The efficacy of the proposed scheme is demonstrated by the finding of 
uncontrolled four-qubit networks that implement either the Toffoli gate,
the Fredkin gate, or remote logic operations. 
The proposed Toffoli gate is stable against imperfections, 
has a high-fidelity for fault
tolerant quantum computation, and is fast, being based on the
non-equilibrium dynamics. 
\end{abstract}

\maketitle

\section*{Introduction}
Computational devices based on the laws of quantum mechanics hold promise to
speed up many algorithms known to be hard for classical computers 
\cite{nielsen2010quantum}. 
The implementation of a full scale computation with existing technology 
requires an outstanding ability
to maintain quantum coherence (i.e. isolation from the environment) without
compromising the ability to control the interactions among the qubits in a
scalable way. 
Among the most successful paradigms of quantum computation, there is the 
``circuit model'', where the algorithm is decomposed into an universal set of 
single- and two-qubit gates \cite{barenco1995elementary}, 
and, to some extent, the so-called adiabatic
quantum computation (AQC) \cite{aharonov2008adiabatic}  
where the output of the algorithm is encoded in the ground
state of an interacting many-qubit Hamiltonian. 
A different approach \cite{benjamin2003quantum} 
is based on the use of always-on interactions, naturally 
occurring between physical qubits, to accomplish the computation.
Compared to the circuit model, this scheme has the advantage of requiring
minimal external control and avoiding the
continuous  switch off and on of 
the interactions between all but two qubits; while compared to AQC it has the
advantage of being faster, being based on the non-equilibrium evolution of the
system.  
Quantum computation with always-on interactions is 
accomplished by combining the natural couplings
with a moderate external
control, e.g. with a smooth shifting of Zeeman energies 
\cite{benjamin2004quantum},  via feedforward techniques 
\cite{satoh2015scalable}, using measurement based computation 
\cite{li2011thermal} or quantum control 
\cite{burgarth2010scalable,muller2011optimizing}.
Most of these approaches are based on the assumption that the natural
couplings are fixed by nature and not tunable, while local interactions can be
modulated with external fields. 
However, the amount of external control required can be minimized 
if the couplings between the qubits can be statically tuned 
\cite{banchi2011nonperturbative}, e.g. during the creation of the quantum 
device. 

The recent advances in the fabrication of superconducting 
quantum devices has opened up to the realization of interacting quantum
networks. In a superconducting device, the qubits are built with a Josephson tunnel
element, an inductance and a capacitor \cite{devoret2013superconducting}, while
local operations and measurements are performed by coupling the qubit
to a resonator 
\cite{wallraff2004strong}. 
The interactions can be designed using lithographic techniques 
by jointly coupling two qubits via 
a capacitor \cite{barends2013coherent} or an inductance \cite{chen2014qubit},
and can be modeled via an effective two-body Hamiltonian 
${\sum_\alpha} J_\alpha \sigma_\alpha{\otimes}\sigma_\alpha$ 
\cite{geller2014tunable,neeley2010generation}, where $\sigma_\alpha$ are the 
Pauli matrices. 
Because of the flexibility in wiring the pairwise interactions among the qubits, it
is possible to arrange them in a planar graph structure, namely a collection of
vertices and links, where the vertices
correspond to the qubits and the links correspond to the 2-body interactions
between them. Moreover, thanks to the development of three-dimensional
superconducting circuits \cite{paik2011observation}, it may be possible in the
near future to wire also non-planar configurations, namely a general qubit 
{\it network}.

Motivated by the above arguments, 
we consider the question whether it is possible to
encode a quantum algorithm into the unmodulated dynamics of a suitably 
large quantum network of pairwise interacting qubits. 
This would be extremely interesting, as it would enable quantum computation 
by simply ``waiting'', without the need of continuously applying external 
control pulses or measurements. 
Even when sequential operations cannot be avoided, our scheme can enable the
{\it in-hardware }
\begin{figure*}[t]
  \centering
  \includegraphics[width=0.85\textwidth]{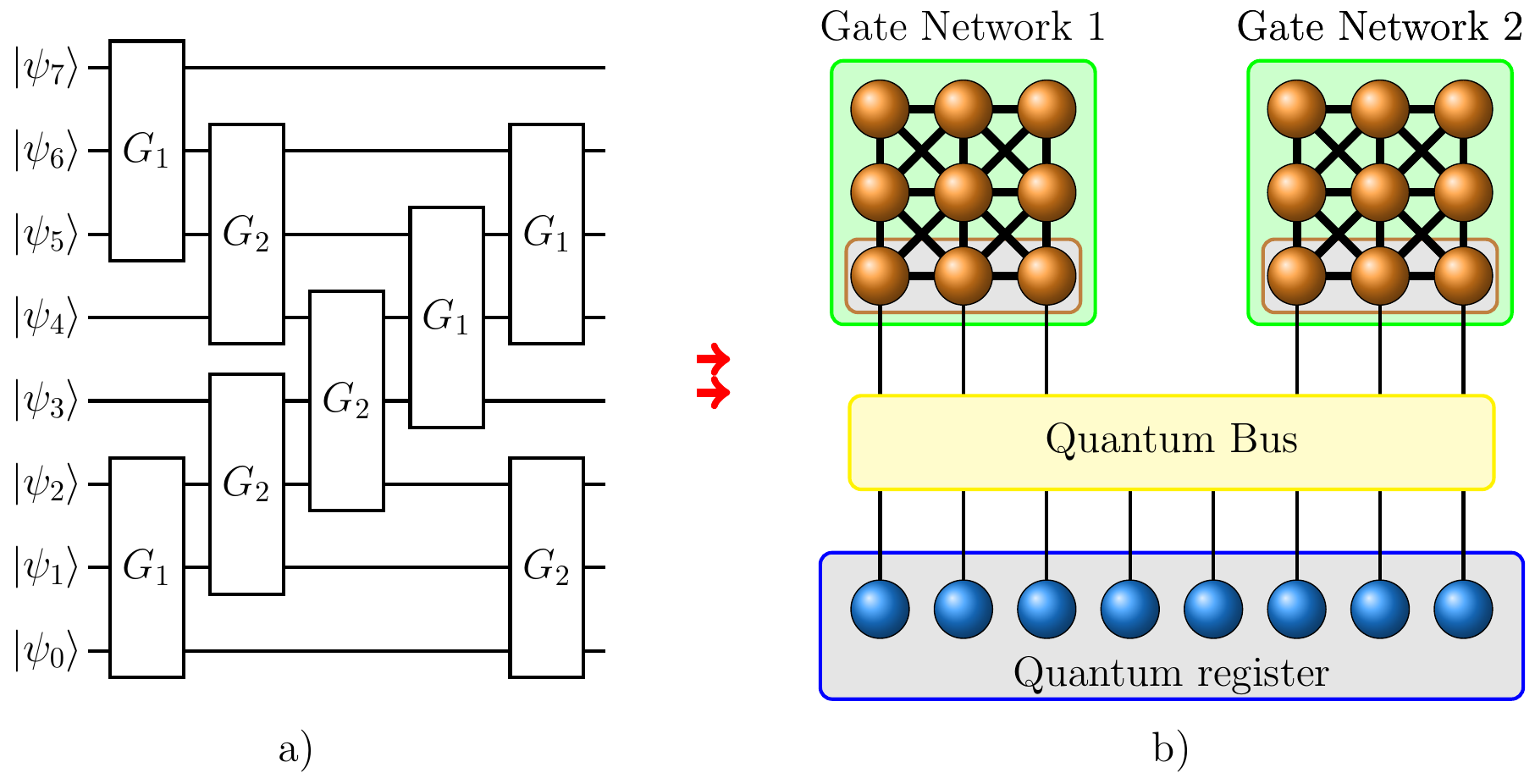}
  \caption{ Schematic transposition of a
      quantum circuit to a trained quantum network. 
      (a) An example 7-qubit circuit where the gates $G_1$ and $G_2$ are 
      sequentially applied. Many important circuits can be cast into the 
      scheme (a), such as those for quantum arithmetics 
      \cite{vedral1996quantum}.  (b) Quantum network implementation of (a):
      each qubit network in the green boxes 
      implements either $G_1$ or $G_2$ on the input/output qubits (the three
      qubits in the bottom row). The quantum bus sequentially transfers the 
      state of three qubits from the register, 
      upon which the gate $G_1$ or $G_2$ has to 
      be applied, to the input/output qubits of the gate network. 
      After the gate
      network has implemented its transformations, the state is transferred 
      back to the original three-qubits of the register.
  }
  \label{fig:circuit}
\end{figure*}
implementation of recurring multi-qubit operations of a quantum algorithm
(see e.g. Fig.\ref{fig:circuit}), such as quantum arithmetic operations
\cite{vedral1996quantum}, and possibly also the quantum Fourier transform
or error correcting codes \cite{nielsen2010quantum}.
We focus on two-body interactions, since they are the most common in 
physical setups, and 
we consider an enlarged network where auxiliary qubits enrich 
the quantum dynamics. 
The important question analyzed in this paper is the
following: given a target unitary operation $U_Q$ on
a given set of qubits $Q$, we consider an extended network $Q{\cup}A$
where $A$ is a set of auxiliary qubits ({\it ancillae}), 
and we ask whether it is possible 
to engineer the pairwise interactions  in $Q{\cup}A$, modeled by the
time-independent Hamiltonian $H_{QA}$, such that
$e^{it H_{QA}}{=}U_Q{\otimes}V_A$ after some time $t$ ($V_A$ may be an extra
unitary operation on the auxiliary space). 
More generally the target operation can depend also on the {\it ancillae} 
initial state:  if  
$e^{it H_{QA}}{=}\sum_n
U^{(n)}_Q{\otimes} 
{\ket{A_n}}{\bra{A_n}}$,  
where $\{\ket{A_n}\}$ form a basis of the {\it ancillae} Hilbert space and, 
e.g., $U_Q{=} U^{(1)}_Q$, then the target operation is implemented when $A$ is
initialized in $\ket{A_1}$.
Our method is particularly useful for implementing quantum gates which requires
$k$-body interactions ($k{>}2$), such as the Toffoli or Fredkin gates
\cite{fedorov2011implementation,cory1998experimental,nielsen2010quantum} 
where $U_Q{\neq}e^{it H_{Q}}$ for any 2-local $H_Q$, 
and for {\it remote logic}, namely for applying a gate to qubits which are
not directly connected but are rather interacting via intermediate systems. 
Our approach is completely different from the simulation of 
$k$-local Hamiltonians with pairwise interactions discussed in the
AQC literature \cite{bravyi2008quantum,biamonte2008realizable}, being based
on the unmodulated dynamics. 
Moreover, being based on unmodulated (time-independent) interactions and 
ancillary qubits, it is
significantly different from quantum optimal control \cite{brif2010control}. 

Our quantum network design procedure is inspired by supervised learning in
feedforward networks \cite{bishop2006pattern}, where the training procedure
involves the optimization of the network couplings (i.e. the {\it weights} 
between different nodes) such that the output
corresponding to some input data has a desired functional form (e.g. for
data classification). 
Although there are many recent developments about using a quantum device
to speed-up machine learning algorithms
\cite{wittek2014quantum,rebentrost2014quantum,paparo2014quantum,wiebe2014quantum,lloyd2014quantum} or storing data 
\cite{rotondo2015dicke}, our optimization procedure is
entirely classical, but specifically developed for quantum hardware {\it
design}. 
Our scheme is completely different from other recent proposals
\cite{nagaj2012universal,bang2008quantum,gammelmark2009quantum} 
because it avoids measurements or 
active feedbacks and requires minimal external control.

\section*{Results}

\subsection*{Supervised quantum network design}
Supervised learning is all about function approximation: given a training set
$\{(I_1, O_1),(I_2,O_2),\dots\}$, namely a collection of inputs $I_k$ and the
corresponding known outputs $O_k$, 
the goal is to find a function $f$ with two desired properties: i)
$O_k {\simeq} f(I_k)$ for any training pair; ii) $f$ should be 
able to infer the unknown output of an input not contained in the training set. 
In classical feedforward networks, the function $f$ is approximated with a 
directed graph organized in layers, where the first layer is the input register
and the last one encodes the output. The value $s_k^{(\ell)}$ of the
$k$-th node in layer $\ell$ is updated via the equation
$  s^{(\ell)}_k {=} A^{\ell}\left[\sum_j
\lambda_{kj}^{(\ell{-}1)}
s^{(\ell{-}1)}_j\right]~,$
where $A^{\ell}$ is an appropriate (typically non-linear) activation function
and $\lambda_{kj}^{(\ell{-}1)}$ is the weight between node $k$ in layer $\ell$
and node $j$ in $\ell{-}1$. 
The training procedure consists in finding the optimal weights
$\V\lambda$ by minimizing a suitable {\it cost function} such as
$\mathcal C{=}\sum_k|O_k{-}f(I_k)|^2$. 

A {\it quantum network}
consists on the other hand of an undirected graph $(V,E)$ of vertices $V$ and links $E$
described by a 2-local Hamiltonian
\begin{align}
  \mathcal H = \sum_{(n,m)\in E} \sum_{\alpha,\beta} J^{\alpha\beta}_{nm}
  \frac{\sigma_n^\alpha \sigma_m^\beta}{4}
  + \sum_{n\in V} \sum_{\alpha} h^{\alpha}_{n} \frac{\sigma_n^\alpha}2~,
  \label{e.Ham}
\end{align}
where $\sigma_n^\alpha$, $\alpha{=}x,y,z$, 
are the Pauli matrices acting on qubit $n$ and, to
simplify the notation, we call $\V\lambda{=}\{J^{\alpha\beta}_{nm},
h^{\alpha}_{n}\}$ the set of parameters. The vertices are composed of
two disjoints sets $V{=}Q{\cup}A$ where $Q$ consists of register qubits 
and $A$ of auxiliary qubits. 
Given a separable initial state $\ket{\psi_Q}{\otimes}\ket{\psi_A}$, the time
evolution according to Hamiltonian \eqref{e.Ham} generates a {\it 
quantum channel} \cite{nielsen2010quantum} $\mathcal E_{\V\lambda}[
\ket{\psi_Q}\bra{\psi_Q}]{=}\Tr_A[e^{-i\mathcal H\tilde t}\ket{\psi_Q}\bra{\psi_Q}
{\otimes}\ket{\psi_A}\bra{\psi_A}e^{i\mathcal H\tilde t}]$ on subsystem $Q$ --
since we
are interested in a fixed operational time $\tilde t$ for simplicity we set
$\tilde t{=}1$, reabsorbing $\tilde t$ 
into the definition of the definition of $\mathcal H$. 
Depending on the flexibility of the experimental apparatus in reliably
initializing the auxiliary qubits, one can add $\ket{\psi_A}$ 
to the set $\V\lambda$. 
Network design consists in the following procedure: 
given a target unitary operation $U_Q$ that we want to implement, 
the goal is to find the parameters $\V\lambda$, if they exist, such that
$\mathcal E_{\V\lambda}[\rho_Q]{=}U_Q\rho_QU_Q^\dagger$ for any $\rho_Q$. 
To simplify the notation we assume that the gate output is encoded in 
$Q$ but it is
straightforward to generalize the formalism when the output sites differ 
from the input ones. 

Motivated by the similarity with classical supervised learning, where
the weights $\V\lambda$ are tuned to 
maximize the ability of the network to reproduce a known output given 
the corresponding input, 
we create a training set $\mathcal T$ with a random set of initial input
states. For each input $\ket\psi{\in}\mathcal T$ the expected known output is 
$U_Q\ket\psi$, while the output of the network evolution is 
$\mathcal E_{\V\lambda}[\ket\psi\bra\psi]$. 
The ``learning'' procedure involves the minimization of the difference between
the output of the network and the expected output, and corresponds to the
maximization of the fidelity 
\begin{align}
  \mathcal F &= \sum_{\ket\psi\in\mathcal T} \frac{\mathcal F_\psi}{|\mathcal T|}, &
  \mathcal F_\psi &= \bra{\psi}U_Q^\dagger\mathcal E_{\V
  \lambda}\left[\ket{\psi}\bra{\psi}\right]U_Q\ket{\psi}.
  \label{e.F}
\end{align}
If the average is performed over all possible states, then  
Eq.\eqref{e.F} can be
substituted by the average gate fidelity 
$\bar{\mathcal F} {=} \int \mathcal F_\psi d\psi$ where the formal 
integration can be explicitly evaluated
\cite{banchi2011nonperturbative,magesan2011gate,pedersen2008distribution}
yielding 
\begin{equation}\label{e.Fave}
  \bar{\mathcal{F}}=\frac1{D+1} + \frac{1}{D(D+1)}
  \sum_{ijkl}U_{ik}^* \mathcal{E}^{ij,kl}_{\V\lambda} U_{jl}~,
\end{equation} 
where
$\mathcal{E}^{ij,kl}_{\V\lambda}{=}\langle{q_i}|\mathcal{E}_{\V\lambda}
\big[|{q_k}\rangle\langle{q_l}|\big]|{q_j}\rangle$, 
$U_{ij}{=}\bra{q_i}U_Q\ket{q_j}$ and $\{\ket{q_j}\}$ form the computational
basis of the $D$-dimensional Hilbert space of qubits $Q$. 
The typical value of the fidelity for 
a random non-optimal evolution of the qubit network
is  $\bar{\mathcal{F}}{=}D^{-1}$, obtained using Haar 
integration techniques \cite{collins2006integration}. This value 
is independent on the details of the {\it ancillae}, since it depends only 
on the dimension of the target Hilbert space, and 
provides an estimate for the initial fidelity of an untrained network.

The gate learning procedure corresponds to a global maximization of the 
fidelity \eqref{e.Fave}. 
However, because of the many parameters in the
Hamiltonian \eqref{e.Ham}, $\bar{\mathcal F}$ can have many local maxima making 
the global optimization extremely complicated. 
Since most global optimization algorithms introduce stochastic strategies, 
rather than introducing unphysical random jumps,  we 
take advantage of the explicit stochastic nature of the problem
($\bar{\mathcal F}$ is a uniform average over random states) and 
we propose the following learning algorithm to design the interactions of
the quantum network. 

\begin{algorithmic}[1]
  \State Choose an initial parameter set $\V\lambda$ (e.g. at random), 
  and choose an initial learning rate $\epsilon$;
  \REPEAT
  \State generate a random $\ket\psi$;
  \State update $L$ times the coupling strengths as
  \begin{align}
    \V\lambda\to \V\lambda  + \epsilon\nabla_{\V\lambda}\bra\psi U_Q^\dagger\mathcal E_{\V
    \lambda}\left[\ket\psi\bra\psi\right]U_Q\ket\psi;
    \label{e.grad}
  \end{align}
  \State decrease $\epsilon$ (see Methods);
  \UNTIL{convergence (or maximum number of operations). }
\end{algorithmic}

Specifically, we combine the above algorithm with the maximization of the
average fidelity (see below) and we observe a drastic speedup of the 
optimization process. 
The parameter $L$ tunes the number of deterministic steps in the learning
procedure, and can be set to the minimum value $L{=}1$, so that after each
interaction the state is changed, or to a higher value. In our simulations
we use $L{=}1$, for simplicity. 
Our algorithm is an application of the stochastic gradient descent 
(SGD) method 
\cite{bottou1998online} to the maximization of the function 
\eqref{e.F}. 
In classical feedforward networks, SGD 
is the {\it de facto} standard algorithm for network training 
\cite{bottou1998online,bishop2006pattern} and is specifically used for large
training datasets, when the evaluation of the cost function and its gradient 
are computationally intensive. 
On the other hand, the average in Eq.\eqref{e.F} can be evaluated explicitly 
over a uniform distribution of an infinite number of initial states,  
giving Eq.\eqref{e.Fave}. 
Although $\mathcal F_\psi$ is easier than $\bar{\mathcal F}$ to compute, 
the major advantage of SGD for quantum network design comes from 
its ability to escape local maxima.  
The crucial observation to show the latter point is that 
the statistical variance over random states ${\rm Var}\mathcal F{=}
\overline{\mathcal F^2}{-}\bar{\mathcal F}^2$ vanishes when $\bar{\mathcal F}{=}1$
(see e.g. \cite{magesan2011gate,pedersen2008distribution}) -- 
indeed, intuitively, since both
$\bar{\mathcal F}$ and $\mathcal F_\psi$ are bounded in $[0,1]$, $\bar{\mathcal
F}$ can achieve its maximum only if $\mathcal F_\psi{=}1$ for all the
states, apart from a set of measure zero. 
On the other hand, if $0{<}\bar{\mathcal F}{<}1$,  then ${\rm Var}\mathcal F{>}0$ and the
fluctuations can be so high that a local maximum of $\bar{\mathcal F}$ may not correspond to a maximum of $\mathcal F_\psi$ for some state $\psi$. 
\begin{figure}[t]
  \centering
  \includegraphics[width=0.45\textwidth]{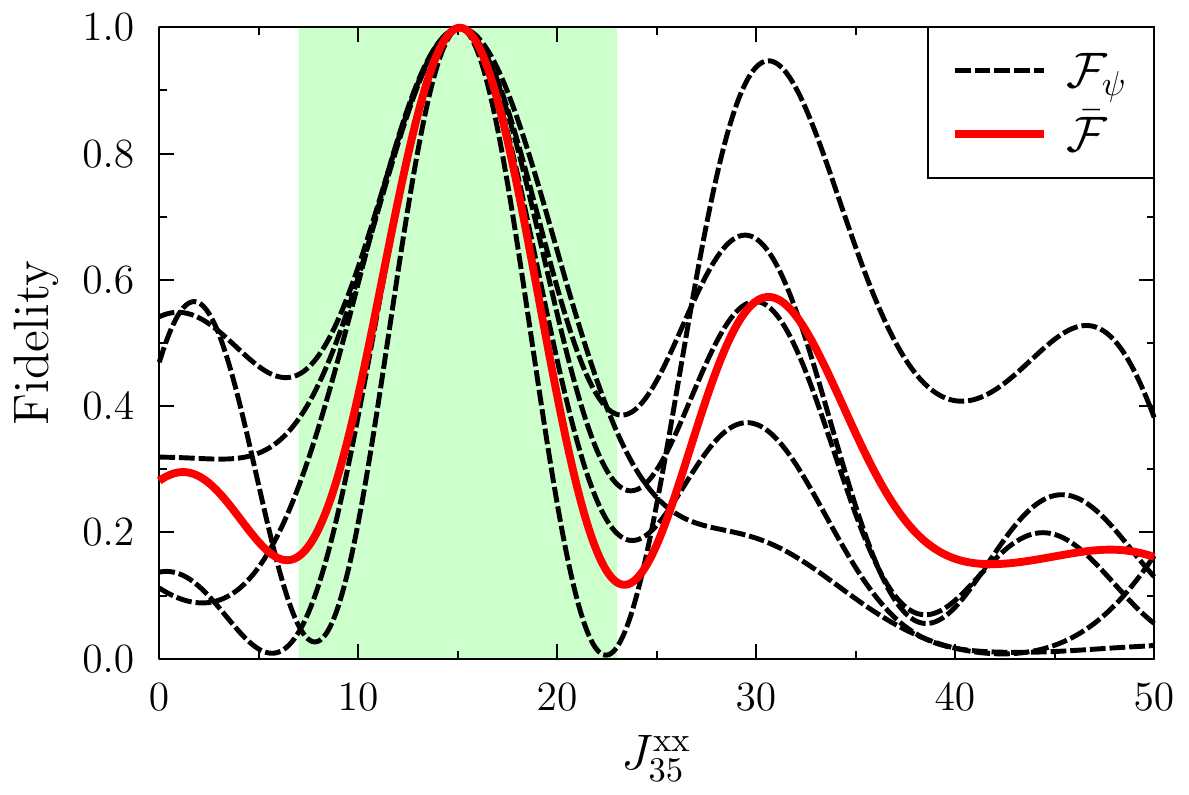}
  \caption{
    Average fidelity $\bar{\mathcal F}$ and fidelity $\mathcal F_\psi$
  for some random states $\psi$ for implementing the Toffoli gate in a 4 qubit
  network (see discussion in the text). All the parameters are set to the
  optimal ones, except $J_{35}^{xx}$ in the abscissa.
  The region around the only global peak is filled in green. 
  }
  \label{fig:toffoli}
\end{figure}
This is indeed shown in Fig.\ref{fig:toffoli} with a real example for the
implementation of the Toffoli gate (see the application section below).
In Fig.\ref{fig:toffoli} the average fidelity
$\bar{\mathcal F}$ has three local maxima
at ${\V \lambda}^{\rm loc.}_k$ ($k=1,2,3$)
and a single global maximum
at ${\V \lambda}^{\rm gl.}$, namely the optimal parameters,
 while the fidelities $\mathcal F_\psi$ for
different random states $\psi$ have a more complicated behavior. 
In view of the argument discussed above, 
all the state fidelities $\mathcal F_\psi$ have 
a global maximum at ${\V \lambda}^{\rm gl.}$ while, remarkably, 
at least one fidelity $\mathcal F_\psi$ has no local maximum at 
${\V \lambda}^{\rm loc.}_k$. 
Our stochastic learning algorithm uses a gradient descent technique for locally
maximising the function $\mathcal F_\psi(\V\lambda)$. 
Therefore, if we are around the slopes of a local
maximum of $\mathcal F_\psi(\V\lambda)$ (say ${\V \lambda}^{\rm loc.}_k$ 
from the previous example) 
and the state $\ket\psi$ is randomly changed
to $\ket\phi$, that local maximum may
disappear from $\mathcal F_\phi(\V\lambda)$ allowing the algorithm to escape
from this non-optimal region when the parameters are updated via
Eq.\eqref{e.grad}. On the other hand, when the algorithm 
is probing the neighborhood of a true optimal point 
 for which $\bar{\mathcal F}(\V\lambda){=}1$, 
(e.g. ${\V \lambda}^{\rm gl.}$ in the previous example), then
the maximum of $\mathcal F_\psi(\V\lambda)$ 
does not disappear when the state $\psi$ is changed, allowing the
``climbing'' procedure to continue.

The above stochastic algorithm may be combined with a deterministic 
maximization
of Eq.\eqref{e.Fave}. In our simulations we use stochastic learning 
for the initial global span of the parameter manifold and, if it
reaches a suitably high fidelity (e.g. $\bar{\mathcal F}{>}95\%$), then it is
reasonable to suppose that the algorithm has found a global maximum. 
Starting from this point we perform a local
maximization of Eq.\eqref{e.Fave} and, if $\bar{\mathcal F}{\simeq}1$ 
is reached, the
learning has been successful. Otherwise we repeat the procedure. 

It is worth emphasizing that given a target gate $U$, it is an open
question to understand {\it a priori} whether a solution may exist for a graph 
with a certain set of interactions (e.g. Heisemberg, Ising, etc.). 
Unlike in quantum control, where
given a time {\it dependent} Hamiltonian $\mathcal H(t)=\mathcal
H_0+\beta(t)\mathcal V$ one can check in advance whether 
$U=T[\exp({-}i\int_0^1 \mathcal H(t)dt)]$ for some control profile $\beta(t)$: 
such profile can exist only if $U$ is contained in the group associated to 
the algebra generated by the repeated commutators of $\mathcal H_0$ and
$\mathcal V$. 
Although no complete algebraic 
characterization is known for our case (see however the 
Methods for a necessary condition) and we have to study each problem 
numerically, in the next sections we find some structures which 
enable the implementation of important quantum gates. 
All numerical simulations have been obtained in a laptop computer using 
QuTiP \cite{Johansson20131234}.

\subsection*{ Application: Toffoli gate}
The Toffoli gate is a key component for many important quantum algorithms,
notably the Shor algorithm \cite{shor1997polynomial}, 
quantum error correction \cite{cory1998experimental}, fault-tolerant 
computation \cite{dennis2001toward}, quantum arithmetic operations
\cite{vedral1996quantum} 
and, together with the Hadamard gate, is universal for quantum computation
\cite{shi2003both}. 
Experimental implementations of this gate has been obtained with
trapped ions \cite{monz2009realization}, superconducting circuits 
\cite{reed2012realization,fedorov2011implementation}, or photonic 
architectures \cite{lanyon2009simplifying}. 
Toffoli gate is a controlled-controlled-not (CCNOT) operation acting on three
qubits. 
It can be implemented in a circuit using five two-qubit
gates \cite{nielsen2010quantum}, or can be obtained in coupled systems via
quantum control techniques \cite{stojanovic2012quantum,zahedinejad2015high}. 
Efficient schemes require higher dimensional system (i.e. qudits) 
\cite{lanyon2009simplifying}. 
On the other hand, the direct implementation using natural interactions is
complicated, since the Hamiltonian $\mathcal H_{\rm CCNOT}$ corresponding to the
gate, i.e. ${\rm CCNOT}{=}e^{i \mathcal H_{\rm CCNOT}}$, has three-body
interactions which unlikely appear in nature. 

By applying our quantum hardware design procedure, we 
show that the Toffoli gate can be implemented in a four qubit network using
only pairwise interactions and constant control fields. 
Our findings enable the construction of a device which implements  
the Toffoli gate with a fidelity 
$\bar{\mathcal F}{=}99.98\%$ by simply
``waiting'' for the natural dynamics to occur, without the need of external
control pulses. 
\begin{figure}[t]
  \centering
  \includegraphics[width=0.4\textwidth]{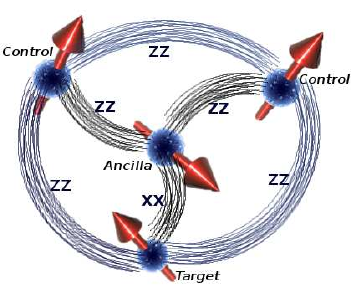}
  \caption{Network implementing the Toffoli gate. The gate acts on the three
  external qubits (the top ones being the control qubits, and the bottom one
being the target), and has an additional auxiliary qubit in the center. }
  \label{fig:graph}
\end{figure}
We consider a four qubit network as displayed in Fig.~\ref{fig:graph}, where the
control qubits are labeled by the indices 1,2, the target is qubit 3 and the
{\it ancilla} is qubit 4. We start our analysis by considering a
fully-connected graph where each qubit interacts with the others 
using XX- and ZZ-type  pairwise interactions, as this kind of interaction
can be
obtained in superconducting circuits \cite{geller2014tunable}. Because of the
symmetries of the Toffoli gate (see Methods), 
we consider the two control qubits to be equally
coupled to the target and the ancilla: 
$J^{\alpha\beta}_{1m}{=}J^{\alpha\beta}_{2m}$, for $m{=}3,4$ and similarly we
set $h^\alpha_1{=}h^\alpha_2$. 
Moreover, since the Toffoli gate is real, we only consider local fields in the
X and Z directions and set $\ket{\psi_A}{=}\cos\eta\ket{\uparrow}{+}
e^{i\xi} \sin\eta \ket{\downarrow}$.
By combining SGD with the maximization of Eq.\eqref{e.Fave} we find the
following optimal parameters, 
\begin{align}
  \nonumber
  J^{zz}_{12} &= -8.940, &
  J^{zz}_{13} &= -4.957, &
  J^{zz}_{14} &= -5.657, \\
  \nonumber
  h^z_1 &= -2.428, &h^z_3 &= J^{zz}_{13}, & h^z_4 &= -0.165, \\
  \nonumber
  h^x_3 &= -19.08,&
  h^x_4 &= -4.267, & J^{xx}_{34} &= 15.06, \\
    \eta &= 0.8182, &\xi & =0.0587,
  \label{e.Jtof}
\end{align}
where the other XX- and ZZ-type interactions not displayed in 
\eqref{e.Jtof} are found to be zero by the learning algorithm, so 
the optimal configuration is the one summarized in Fig.\ref{fig:graph}
where the XX coupling is only between qubits 3 and 4. 
In more physical terms, if the maximal allowed
coupling is fixed to $J/2\pi{\approx}40 {\rm MHz}$, then we find a gate time of
$60 {\rm ns}$ and 
\begin{align}
  \nonumber
  J^{zz}_{12} &= -149.2 \,{\rm MHz}, &
  J^{zz}_{13} &= -82.71 \,{\rm MHz}, 
  \nonumber \\ 
  J^{zz}_{14} &= -94.39 \,{\rm MHz}, & 
  J^{xx}_{34} &= 251.3\, {\rm MHz}, 
  \nonumber \\ 
  h^x_3 &= -318.4\, {\rm MHz},&
  h^x_4 &= -71.2\, {\rm MHz}, 
  \nonumber \\
  h^z_1 &= -40.52 \,{\rm MHz}, 
  & h^z_4 &= -2.751\, {\rm MHz}, 
  \nonumber \\
  h^z_3 &= J^{zz}_{13}.
  \label{e.JtofMHz}
\end{align}
With the optimal parameters of Eqs.\eqref{e.Jtof},\eqref{e.JtofMHz} 
we obtain an average gate fidelity of 99.98\%, above the threshold for
topological fault tolerance for single- and two-qubit gates, while 
by avoiding the extra phase fixing
$\xi{=}0$ we still obtain $\bar{\mathcal F}{=}99.92\%$. 
Moreover, our gate fidelity is above the Toffoli gate accuracy 
threshold ($755/756{\simeq}99.87\%$) for fault-tolerant computation
in the limit in which Clifford gate errors are negligible 
\cite{gaitan2008quantum}.

The optimal parameters \eqref{e.Jtof},\eqref{e.JtofMHz}  
are stable against an imperfect
tuning of the interactions. Indeed, we considered a perturbation
$\lambda_k{\to}\lambda_k{+}\epsilon r_k$,  $r_k{\in}[0,1]$ being a random
number and $\epsilon$ being the strength of the static perturbation, and
found that 
$\bar{\mathcal F}{>}99.9\%$ if $\epsilon{<}0.04$ ($\epsilon{<}0.7\,{\rm MHz})$
and 
$\bar{\mathcal F}{>}99\%$ if $\epsilon{<}0.18$ ($\epsilon{<}3\,{\rm MHz})$.

\subsection*{Application: Fredkin gate}
Fredkin gate is a controlled-swap (CSWAP) operation acting on three qubits 
which is universal for reversible computation \cite{nielsen2010quantum}. 
We found
that this gate can be obtained with perfect fidelity (up to the numerical
precision)
in a four qubit network with Hamiltonian 
\eqref{e.Ham} where
$J^{xx}_{12}{=} J^{xx}_{13}{=}13.60$ $(227.0\, {\rm MHz})$,
$J^{\alpha\alpha}_{23}{=}{-}4.712$ $(-78.62\, {\rm MHz})$,
$J^{xx}_{24} {=} J^{xx}_{34} {=} 8.400$ $(140.2\, {\rm MHz})$, 
$J^{zz}_{12} {=} J^{zz}_{13} {=} 11.15$ $(186.1 \,{\rm MHz})$, 
$h^x_4{=} 1.025$ $(17.11 \, {\rm MHz})$, 
$h^z_1 {=}\pi$ $(54.42\, {\rm MHz})$. 
The values in MHz correspond to a gate time of 60ns.
Moreover, $e^{ {-}i\mathcal H}{=}{\rm CSWAP}_{123}{\otimes}U_4$ so the gate is
independent on the initial state of the ancilla.
As for the Toffoli gate, this optimal configuration has been obtained by
starting the training procedure with a fully connected graph with all the 
interactions, so the fact that some interactions are zero is a result of 
the optimization process. 

\subsection*{Application: remote logic}
We study a qubit network which implements a maximally entangling gate
between two sites which are not directly coupled.  
Remote logic has been studied extensively in spin chains 
for achieving entangling operations between the boundary sites 
\cite{banchi2011nonperturbative,yao2013quantum,banchi2015perfect,benjamin2003quantum},
and it is a 
building block for a proposed architecture for solid-state quantum computation
at room temperature \cite{yao2012scalable}. 
For simplicity we consider a SU(2) invariant four qubit network, 
interacting with a Heisenberg Hamiltonian
$H{=}\sum_{i{\neq}j=1}^4\sum_{\alpha{=}x,y,z}
J_{ij}\sigma_i^{\alpha}\sigma_j^\alpha/4$ where there is
no direct coupling between qubits 1 and 4 ($J_{14}{=}0$). 
Applying our learning algorithm we found that the 
$\sqrt{\rm SWAP}$ gate, which is universal for quantum computation when paired
with single qubit operations \cite{nielsen2010quantum}, 
can be achieved between qubits 1 and 4 with unit fidelity with different
choices of 
$J_{12}{=}J_{24}$,
$J_{13}{=}J_{34}$, and 
$J_{23}$ when  the initial state of ancillae is
$(\ket{\uparrow\downarrow}{-}\ket{\downarrow\uparrow})/\sqrt2$. 
Given this simplification one can then find a solution analytically: 
$J_{12}{=}\alpha{+}\pi\sqrt{(2n)^2{-}1}{/}\sqrt{8}$,
$J_{13}{=}\alpha{-}\pi\sqrt{(2n)^2{-}1}{/}\sqrt{8}$,
$J_{23}{=}\alpha+(-1)^n\pi$, where $n$ is an integer. 
We find analytically that irrespective of $\alpha$ the above choice gives
perfect fidelity. 
Our strategy has not found any three-qubit configurations 
which implement a remote 
$\sqrt{\rm SWAP}$ gate, so the four qubit network is the minimal non-trivial
example. 
Remarkably, some of our 4-qubit configuration are 
more stable to noise than the direct
implementation of the gate in a two-qubit system (namely when $J_{14}{=}\pi/2$
and the other couplings are zero). For instance, if 
$J_{ij}=J_{ij}^{\rm optimal}+\epsilon$, $J_{ij}^{\rm optimal}$ being the optimal
value for implementing the gate, we found that when $\epsilon$ is
randomly distributed in $[0,1/2]$, then the 4-qubit system with $n=1$ 
still has, on average, $\bar{\mathcal F}\simeq 99.1\%$, while the 
direct two-qubit case has 
 $\bar{\mathcal F}\simeq 98.8\%$.

\subsection*{Towards a scalable architecture for quantum computation}
Current architectures for quantum computation, e.g. 
with superconducting qubits \cite{barends2014superconducting} or
ion traps \cite{Monz1068}, are based on an arrays of interacting 
qubits which are continuously controlled via external pulses to 
implement the desired operation. This approach may suffer from scalability
issues because, even assuming the ability to maintain quantum coherence for
a long time,
extremely large (classical) control units will be necessary
to generate the sophisticated pulse sequences required to implement a 
full-scale quantum algorithm.
On the other hand, the approach that we have in mind shares more similarities
with integrated circuits in nowadays electronics, 
where a set of special-purpose logic units (modules) 
are wired together to achieve computation (or other tasks). 
In our vision, different modules can be fabricated with qubit networks 
designed to 
produce a specific logic task, namely a quantum gate, automatically 
without the need of external control. As in
Fig.\ref{fig:circuit} the different logic and memory units 
can be reciprocally connected using a 
quantum bus, whose purpose is to transfer the qubit states between the quantum
registers/memory and the input/output qubits of the modules. 
In Fig.\ref{fig:circuit} for simplicity the input/output ports of the 
modules are designed in the same physical 
qubits, although this can be easily extended to more general cases. 
The quantum bus can be realized with different technologies, e.g. with 
microwave resonators \cite{mariantoni2011implementing}, or can also 
be implemented via quantum state transfer in a qubit network 
\cite{nikolopoulos2014quantum}.
The modules shown in Fig.\ref{fig:circuit} can be designed to produce 
either simple basic operations, like the CNOT or the Toffoli gate, or, in
principle, they can directly implement larger components of a quantum algorithm
like the Quantum Fourier Transform or error correcting codes 
\cite{nielsen2010quantum}. 
In this respect, to treat systems with many parameters one can easily 
combine our optimization strategy, based on fidelity statistics, with
metaheuristic strategies \cite{storn1997differential} which simultaneously 
deals with many candidate solutions and are known to be fast in global
optimization with high-dimensional parameter spaces. Moreover, 
highly optimized deep learning algorithms are already used to train 
neural networks with 60 millions of parameters \cite{krizhevsky2012imagenet}.
However, 
given the difficulty in numerically simulating large quantum systems, this
approach may be reasonable for networks up to, say, 20-30 qubits.

\section*{Discussion}
Inspired by classical supervised learning, 
we have proposed an optimization scheme 
to encode a quantum operation into the unmodulated dynamics of a qubit
register, which is part of a bigger
network of pairwise interacting qubits. 
Our strategy is based on the static engineering of the pairwise couplings, 
and enables the creation of a quantum device which implements the
desired operation 
by simply waiting for the natural dynamics to occur, without 
the need of external control pulses. 
Our findings show that machine learning inspired techniques 
can be combined with quantum
mechanics not only for data classification speed-up
\cite{rebentrost2014quantum,wittek2014quantum} or
quantum black-box certification \cite{bisio2010optimal,wiebe2014hamiltonian},
but also for quantum hardware design. 

This paper opens up the topic of encoding quantum gates and operations 
into the unmodulated dynamics of qubit networks. 
Although we have focused on small systems, larger networks 
can be considered using more efficient training schemes. 
These would enable the simulation
of larger components of a quantum algorithm, since different multi-qubit gates 
can be combined into a unique quantum operation which can be simulated in a large quantum network. 
Moreover, when combined with a quantum bus as in  Fig. \ref{fig:circuit}, 
our strategy can provide an alternative approach to universal quantum
computation which avoids the decomposition of the algorithm into
one- and two-qubit gates. 
Note that most quantum algorithms take {\it classical} inputs so the extra
control required for initialization demands the further ability to
fully polarise globally the spins. The latter step is
however typically much easier than the implementation of entangling gates, which
has been considered in this paper. 
%
Moreover, in view of the recent 
experimental measures of the average gate fidelity \cite{lu2015experimental},
it is tempting to predict an all-quantum version of our learning procedure
where $\bar{\mathcal F}$ is not classically simulated, but rather directly
measured. This would require a further highly controlled system to infer
the optimal parameters of an uncontrolled quantum network,
which can be used to industrialize the production of unmodulated quantum 
devices implementing the desired algorithm. 

Our results demonstrate the efficacy of the proposed scheme in designing
four qubit networks which implement the
Toffoli and Fredkin gates or remote logic operations. 
The proposed Toffoli gate is fast, has high-fidelity for fault-tolerant
computation, and  only uses static XX- and ZZ-type interactions which 
can be achieved in superconducting systems \cite{geller2014tunable}. 
The key advantage of our method is in exploiting all the permanent 
interactions in the qubit network
without trying to suppress some of them sequentially to implement 
pairwise gates.
Moreover, being based on non-equilibrium
dynamics, our gate is fast: if $J/2\pi{\approx}40 {\rm MHz}$ then the total 
operation time is 
around $60 {\rm ns}$ which matches the current gate times for single and two
qubit operations \cite{barends2014superconducting}.

\appendix
\section*{Methods}
\subsection*{Learning rate} 
The choice of the learning rate $\epsilon$ is crucial. 
If the initial learning rate is too small, it might
not escape from  the different ``local maximum'' points, 
while if it is too large it will continue to
randomly jump without even seeing the local maxima. 
To maximize the speed and precision of SDG the learning rate $\epsilon$ has to 
decrease as a function of the steps, a common choice being
$\epsilon{\propto}m^{-1/2}$ where $m$ is 
the step counter \cite{bottou1998online}. 
However, when the gradient in Eq.\eqref{e.grad} 
cannot be performed analytically,
one can use more sophisticated techniques \cite{spall2005introduction}  where
both the learning rate and the finite difference approximation of the gradient 
change as a function of $m$.

\subsection*{Symmetries}
In the design of the quantum network and its couplings
the number of parameters can be drastically reduced if the 
target unitary operation $U_Q$ has some symmetries, namely if there exists some
unitary matrix $S$ such that $[U_Q,S_Q]{=}0$. This condition requires the
quantum channel 
$\mathcal E_{\V\lambda}[\rho]{=}\Tr_A[e^{-i\mathcal H}\rho
{\otimes}\rho_Ae^{i\mathcal H}]$ 
to satisfy  
$\mathcal E_{\V\lambda}(S_Q\rho S_Q^\dagger){=} 
S_Q\mathcal E_{\V\lambda}(\rho) S_Q^\dagger$ for each state $\rho$, e.g.
$[\mathcal H,S_Q{\otimes}\openone_A]{=}0$. 
Conversely, if the interaction type is fixed by nature 
(for instance, 
only Ising or Heisenberg interactions are
allowed), then one has to check 
whether the Lie algebra spanned by the
operators in $\mathcal H$ contains the generators of $U_Q$. 


\subsection*{Bottom-up construction: Lie algebraic characterization}
All the numerical results presented in the main text are obtained using a
{\it top-down} approach: after selecting the interaction types (e.g. XX, ZZ,
Heisenberg etc.), 
the algorithm starts with a zero-bias fully connected configuration where
all the qubit pairs of the network interacts with all possible interactions,
each weighted with a different parameter, and different local fields. 
As a result of the training procedure, we found numerically that 
most of these parameters are indeed zero.
However, for larger networks it is better to use a {\it bottom-up} 
approach where one starts with a minimal set of parameters, and then adds
other parameters until a solution is found. 

To construct a minimal set of parameters one can use a Lie algebraic
characterization inspired by quantum control. We write the Hamiltonian as 
$\mathcal H{=} \sum_j \lambda_j \mathcal O_j$, where $\lambda_j$ are the 
independent parameters and $\mathcal O_j$ the operators. If the parameters
are time dependent, then there exist suitable pulses $\lambda_j(t)$ such
that the dynamics implements the target gate $G$ only if $\log(G)$ is 
contained in the algebra generated by the repeated commutators
$[\mathcal O_j, [\mathcal O_k, \dots]]$. Since our scheme is based on the 
particular choice where $\lambda_j(t)$ is constant, the above characterization
still provides a necessary condition. 
As an example, we consider the Toffoli gate and the solution Eq.\eqref{e.Jtof}
where 
$\mathcal O_1{=}\sigma_1^z\sigma_2^z$,
$\mathcal O_2{=}\sigma_1^z\sigma_3^z {+}\sigma_2^z\sigma_3^z {+}2\sigma_3^z$,
$\mathcal O_3{=}\sigma_1^z\sigma_4^z {+}\sigma_2^z\sigma_4^z $,
$\mathcal O_4{=}\sigma_3^x\sigma_4^x$,
$\mathcal O_5{=}\sigma_1^z {+}\sigma_2^z $,
$\mathcal O_7{=}\sigma_4^z $,
$\mathcal O_8{=}\sigma_3^x $,
$\mathcal O_9{=}\sigma_4^x $. 
It is simple to check that $\log G$ (up to an irrelevant constant factor) 
is contained in the algebra generated by the operators $O_j$, while this is not
the case if the operator $\mathcal O_8$ is removed from the Hamiltonian. 
Therefore, no solution is possible if $\lambda_8{\equiv}0$. 

Inspired by the above example the bottom-up approach consists in the following
steps: i) based on the symmetries of the target gate and on the physically
allowed interactions one defines an initial set of operators; ii) 
other operators are added to the set until 
the dynamical algebra contains $\log(G)$; iii) one starts the numerical 
parameter training to check for convergence (different runs may be required). 
Until the solution is found one then either adds new operators, or change
the previous ones.

\section*{Acknowledgments}
LB,SB 
acknowledge the financial support by the ERC under Starting Grant 308253
PACOMANEDIA. The authors thank P. Wittek, A. Monras and J.I. Cirac 
for their valuable comments and suggestions.


\begin{thebibliography}{60}%
\makeatletter
\providecommand \@ifxundefined [1]{%
 \@ifx{#1\undefined}
}%
\providecommand \@ifnum [1]{%
 \ifnum #1\expandafter \@firstoftwo
 \else \expandafter \@secondoftwo
 \fi
}%
\providecommand \@ifx [1]{%
 \ifx #1\expandafter \@firstoftwo
 \else \expandafter \@secondoftwo
 \fi
}%
\providecommand \natexlab [1]{#1}%
\providecommand \enquote  [1]{``#1''}%
\providecommand \bibnamefont  [1]{#1}%
\providecommand \bibfnamefont [1]{#1}%
\providecommand \citenamefont [1]{#1}%
\providecommand \href@noop [0]{\@secondoftwo}%
\providecommand \href [0]{\begingroup \@sanitize@url \@href}%
\providecommand \@href[1]{\@@startlink{#1}\@@href}%
\providecommand \@@href[1]{\endgroup#1\@@endlink}%
\providecommand \@sanitize@url [0]{\catcode `\\12\catcode `\$12\catcode
  `\&12\catcode `\#12\catcode `\^12\catcode `\_12\catcode `\%12\relax}%
\providecommand \@@startlink[1]{}%
\providecommand \@@endlink[0]{}%
\providecommand \url  [0]{\begingroup\@sanitize@url \@url }%
\providecommand \@url [1]{\endgroup\@href {#1}{\urlprefix }}%
\providecommand \urlprefix  [0]{URL }%
\providecommand \Eprint [0]{\href }%
\providecommand \doibase [0]{http://dx.doi.org/}%
\providecommand \selectlanguage [0]{\@gobble}%
\providecommand \bibinfo  [0]{\@secondoftwo}%
\providecommand \bibfield  [0]{\@secondoftwo}%
\providecommand \translation [1]{[#1]}%
\providecommand \BibitemOpen [0]{}%
\providecommand \bibitemStop [0]{}%
\providecommand \bibitemNoStop [0]{.\EOS\space}%
\providecommand \EOS [0]{\spacefactor3000\relax}%
\providecommand \BibitemShut  [1]{\csname bibitem#1\endcsname}%
\let\auto@bib@innerbib\@empty
\bibitem [{\citenamefont {Nielsen}\ and\ \citenamefont
  {Chuang}(2000)}]{nielsen2010quantum}%
  \BibitemOpen
  \bibfield  {author} {\bibinfo {author} {\bibfnamefont {M.~A.}\ \bibnamefont
  {Nielsen}}\ and\ \bibinfo {author} {\bibfnamefont {I.~L.}\ \bibnamefont
  {Chuang}},\ }\href@noop {} {\emph {\bibinfo {title} {Quantum computation and
  quantum information}}}\ (\bibinfo  {publisher} {Cambridge University Press},\
  \bibinfo {year} {2000})\BibitemShut {NoStop}%
\bibitem [{\citenamefont {Barenco}\ \emph {et~al.}(1995)\citenamefont {Barenco}
  \emph {et~al.}}]{barenco1995elementary}%
  \BibitemOpen
  \bibfield  {author} {\bibinfo {author} {\bibfnamefont {A.}~\bibnamefont
  {Barenco}} \emph {et~al.},\ }\href@noop {} {\bibfield  {journal} {\bibinfo
  {journal} {Phys. Rev. A}\ }\textbf {\bibinfo {volume} {52}},\ \bibinfo
  {pages} {3457} (\bibinfo {year} {1995})}\BibitemShut {NoStop}%
\bibitem [{\citenamefont {Aharonov}\ \emph {et~al.}(2008)\citenamefont
  {Aharonov}, \citenamefont {Van~Dam}, \citenamefont {Kempe}, \citenamefont
  {Landau}, \citenamefont {Lloyd},\ and\ \citenamefont
  {Regev}}]{aharonov2008adiabatic}%
  \BibitemOpen
  \bibfield  {author} {\bibinfo {author} {\bibfnamefont {D.}~\bibnamefont
  {Aharonov}}, \bibinfo {author} {\bibfnamefont {W.}~\bibnamefont {Van~Dam}},
  \bibinfo {author} {\bibfnamefont {J.}~\bibnamefont {Kempe}}, \bibinfo
  {author} {\bibfnamefont {Z.}~\bibnamefont {Landau}}, \bibinfo {author}
  {\bibfnamefont {S.}~\bibnamefont {Lloyd}}, \ and\ \bibinfo {author}
  {\bibfnamefont {O.}~\bibnamefont {Regev}},\ }\href@noop {} {\bibfield
  {journal} {\bibinfo  {journal} {SIAM review}\ }\textbf {\bibinfo {volume}
  {50}},\ \bibinfo {pages} {755} (\bibinfo {year} {2008})}\BibitemShut
  {NoStop}%
\bibitem [{\citenamefont {Benjamin}\ and\ \citenamefont
  {Bose}(2003)}]{benjamin2003quantum}%
  \BibitemOpen
  \bibfield  {author} {\bibinfo {author} {\bibfnamefont {S.~C.}\ \bibnamefont
  {Benjamin}}\ and\ \bibinfo {author} {\bibfnamefont {S.}~\bibnamefont
  {Bose}},\ }\href@noop {} {\bibfield  {journal} {\bibinfo  {journal} {Physical
  review letters}\ }\textbf {\bibinfo {volume} {90}},\ \bibinfo {pages}
  {247901} (\bibinfo {year} {2003})}\BibitemShut {NoStop}%
\bibitem [{\citenamefont {Benjamin}\ and\ \citenamefont
  {Bose}(2004)}]{benjamin2004quantum}%
  \BibitemOpen
  \bibfield  {author} {\bibinfo {author} {\bibfnamefont {S.~C.}\ \bibnamefont
  {Benjamin}}\ and\ \bibinfo {author} {\bibfnamefont {S.}~\bibnamefont
  {Bose}},\ }\href@noop {} {\bibfield  {journal} {\bibinfo  {journal} {Physical
  Review A}\ }\textbf {\bibinfo {volume} {70}},\ \bibinfo {pages} {032314}
  (\bibinfo {year} {2004})}\BibitemShut {NoStop}%
\bibitem [{\citenamefont {Satoh}\ \emph {et~al.}(2015)\citenamefont {Satoh},
  \citenamefont {Matsuzaki}, \citenamefont {Kakuyanagi}, \citenamefont {Munro},
  \citenamefont {Semba}, \citenamefont {Yamaguchi},\ and\ \citenamefont
  {Saito}}]{satoh2015scalable}%
  \BibitemOpen
  \bibfield  {author} {\bibinfo {author} {\bibfnamefont {T.}~\bibnamefont
  {Satoh}}, \bibinfo {author} {\bibfnamefont {Y.}~\bibnamefont {Matsuzaki}},
  \bibinfo {author} {\bibfnamefont {K.}~\bibnamefont {Kakuyanagi}}, \bibinfo
  {author} {\bibfnamefont {W.~J.}\ \bibnamefont {Munro}}, \bibinfo {author}
  {\bibfnamefont {K.}~\bibnamefont {Semba}}, \bibinfo {author} {\bibfnamefont
  {H.}~\bibnamefont {Yamaguchi}}, \ and\ \bibinfo {author} {\bibfnamefont
  {S.}~\bibnamefont {Saito}},\ }\href@noop {} {\bibfield  {journal} {\bibinfo
  {journal} {Physical Review A}\ }\textbf {\bibinfo {volume} {91}},\ \bibinfo
  {pages} {052329} (\bibinfo {year} {2015})}\BibitemShut {NoStop}%
\bibitem [{\citenamefont {Li}\ \emph {et~al.}(2011)\citenamefont {Li},
  \citenamefont {Browne}, \citenamefont {Kwek}, \citenamefont {Raussendorf},\
  and\ \citenamefont {Wei}}]{li2011thermal}%
  \BibitemOpen
  \bibfield  {author} {\bibinfo {author} {\bibfnamefont {Y.}~\bibnamefont
  {Li}}, \bibinfo {author} {\bibfnamefont {D.~E.}\ \bibnamefont {Browne}},
  \bibinfo {author} {\bibfnamefont {L.~C.}\ \bibnamefont {Kwek}}, \bibinfo
  {author} {\bibfnamefont {R.}~\bibnamefont {Raussendorf}}, \ and\ \bibinfo
  {author} {\bibfnamefont {T.-C.}\ \bibnamefont {Wei}},\ }\href@noop {}
  {\bibfield  {journal} {\bibinfo  {journal} {Physical review letters}\
  }\textbf {\bibinfo {volume} {107}},\ \bibinfo {pages} {060501} (\bibinfo
  {year} {2011})}\BibitemShut {NoStop}%
\bibitem [{\citenamefont {Burgarth}\ \emph {et~al.}(2010)\citenamefont
  {Burgarth}, \citenamefont {Maruyama}, \citenamefont {Murphy}, \citenamefont
  {Montangero}, \citenamefont {Calarco}, \citenamefont {Nori},\ and\
  \citenamefont {Plenio}}]{burgarth2010scalable}%
  \BibitemOpen
  \bibfield  {author} {\bibinfo {author} {\bibfnamefont {D.}~\bibnamefont
  {Burgarth}}, \bibinfo {author} {\bibfnamefont {K.}~\bibnamefont {Maruyama}},
  \bibinfo {author} {\bibfnamefont {M.}~\bibnamefont {Murphy}}, \bibinfo
  {author} {\bibfnamefont {S.}~\bibnamefont {Montangero}}, \bibinfo {author}
  {\bibfnamefont {T.}~\bibnamefont {Calarco}}, \bibinfo {author} {\bibfnamefont
  {F.}~\bibnamefont {Nori}}, \ and\ \bibinfo {author} {\bibfnamefont {M.~B.}\
  \bibnamefont {Plenio}},\ }\href@noop {} {\bibfield  {journal} {\bibinfo
  {journal} {Physical Review A}\ }\textbf {\bibinfo {volume} {81}},\ \bibinfo
  {pages} {040303} (\bibinfo {year} {2010})}\BibitemShut {NoStop}%
\bibitem [{\citenamefont {M{\"u}ller}\ \emph {et~al.}(2011)\citenamefont
  {M{\"u}ller}, \citenamefont {Reich}, \citenamefont {Murphy}, \citenamefont
  {Yuan}, \citenamefont {Vala}, \citenamefont {Whaley}, \citenamefont
  {Calarco},\ and\ \citenamefont {Koch}}]{muller2011optimizing}%
  \BibitemOpen
  \bibfield  {author} {\bibinfo {author} {\bibfnamefont {M.}~\bibnamefont
  {M{\"u}ller}}, \bibinfo {author} {\bibfnamefont {D.}~\bibnamefont {Reich}},
  \bibinfo {author} {\bibfnamefont {M.}~\bibnamefont {Murphy}}, \bibinfo
  {author} {\bibfnamefont {H.}~\bibnamefont {Yuan}}, \bibinfo {author}
  {\bibfnamefont {J.}~\bibnamefont {Vala}}, \bibinfo {author} {\bibfnamefont
  {K.}~\bibnamefont {Whaley}}, \bibinfo {author} {\bibfnamefont
  {T.}~\bibnamefont {Calarco}}, \ and\ \bibinfo {author} {\bibfnamefont
  {C.}~\bibnamefont {Koch}},\ }\href@noop {} {\bibfield  {journal} {\bibinfo
  {journal} {Physical Review A}\ }\textbf {\bibinfo {volume} {84}},\ \bibinfo
  {pages} {042315} (\bibinfo {year} {2011})}\BibitemShut {NoStop}%
\bibitem [{\citenamefont {Banchi}\ \emph {et~al.}(2011)\citenamefont {Banchi},
  \citenamefont {Bayat}, \citenamefont {Verrucchi},\ and\ \citenamefont
  {Bose}}]{banchi2011nonperturbative}%
  \BibitemOpen
  \bibfield  {author} {\bibinfo {author} {\bibfnamefont {L.}~\bibnamefont
  {Banchi}}, \bibinfo {author} {\bibfnamefont {A.}~\bibnamefont {Bayat}},
  \bibinfo {author} {\bibfnamefont {P.}~\bibnamefont {Verrucchi}}, \ and\
  \bibinfo {author} {\bibfnamefont {S.}~\bibnamefont {Bose}},\ }\href@noop {}
  {\bibfield  {journal} {\bibinfo  {journal} {Physical review letters}\
  }\textbf {\bibinfo {volume} {106}},\ \bibinfo {pages} {140501} (\bibinfo
  {year} {2011})}\BibitemShut {NoStop}%
\bibitem [{\citenamefont {Devoret}\ and\ \citenamefont
  {Schoelkopf}(2013)}]{devoret2013superconducting}%
  \BibitemOpen
  \bibfield  {author} {\bibinfo {author} {\bibfnamefont {M.}~\bibnamefont
  {Devoret}}\ and\ \bibinfo {author} {\bibfnamefont {R.}~\bibnamefont
  {Schoelkopf}},\ }\href@noop {} {\bibfield  {journal} {\bibinfo  {journal}
  {Science}\ }\textbf {\bibinfo {volume} {339}},\ \bibinfo {pages} {1169}
  (\bibinfo {year} {2013})}\BibitemShut {NoStop}%
\bibitem [{\citenamefont {Wallraff}\ \emph {et~al.}(2004)\citenamefont
  {Wallraff}, \citenamefont {Schuster}, \citenamefont {Blais}, \citenamefont
  {Frunzio}, \citenamefont {Huang}, \citenamefont {Majer}, \citenamefont
  {Kumar}, \citenamefont {Girvin},\ and\ \citenamefont
  {Schoelkopf}}]{wallraff2004strong}%
  \BibitemOpen
  \bibfield  {author} {\bibinfo {author} {\bibfnamefont {A.}~\bibnamefont
  {Wallraff}}, \bibinfo {author} {\bibfnamefont {D.~I.}\ \bibnamefont
  {Schuster}}, \bibinfo {author} {\bibfnamefont {A.}~\bibnamefont {Blais}},
  \bibinfo {author} {\bibfnamefont {L.}~\bibnamefont {Frunzio}}, \bibinfo
  {author} {\bibfnamefont {R.-S.}\ \bibnamefont {Huang}}, \bibinfo {author}
  {\bibfnamefont {J.}~\bibnamefont {Majer}}, \bibinfo {author} {\bibfnamefont
  {S.}~\bibnamefont {Kumar}}, \bibinfo {author} {\bibfnamefont {S.~M.}\
  \bibnamefont {Girvin}}, \ and\ \bibinfo {author} {\bibfnamefont {R.~J.}\
  \bibnamefont {Schoelkopf}},\ }\href@noop {} {\bibfield  {journal} {\bibinfo
  {journal} {Nature}\ }\textbf {\bibinfo {volume} {431}},\ \bibinfo {pages}
  {162} (\bibinfo {year} {2004})}\BibitemShut {NoStop}%
\bibitem [{\citenamefont {Barends}\ \emph {et~al.}(2013)\citenamefont
  {Barends}, \citenamefont {Kelly}, \citenamefont {Megrant}, \citenamefont
  {Sank}, \citenamefont {Jeffrey}, \citenamefont {Chen}, \citenamefont {Yin},
  \citenamefont {Chiaro}, \citenamefont {Mutus}, \citenamefont {Neill} \emph
  {et~al.}}]{barends2013coherent}%
  \BibitemOpen
  \bibfield  {author} {\bibinfo {author} {\bibfnamefont {R.}~\bibnamefont
  {Barends}}, \bibinfo {author} {\bibfnamefont {J.}~\bibnamefont {Kelly}},
  \bibinfo {author} {\bibfnamefont {A.}~\bibnamefont {Megrant}}, \bibinfo
  {author} {\bibfnamefont {D.}~\bibnamefont {Sank}}, \bibinfo {author}
  {\bibfnamefont {E.}~\bibnamefont {Jeffrey}}, \bibinfo {author} {\bibfnamefont
  {Y.}~\bibnamefont {Chen}}, \bibinfo {author} {\bibfnamefont {Y.}~\bibnamefont
  {Yin}}, \bibinfo {author} {\bibfnamefont {B.}~\bibnamefont {Chiaro}},
  \bibinfo {author} {\bibfnamefont {J.}~\bibnamefont {Mutus}}, \bibinfo
  {author} {\bibfnamefont {C.}~\bibnamefont {Neill}},  \emph {et~al.},\
  }\href@noop {} {\bibfield  {journal} {\bibinfo  {journal} {Physical review
  letters}\ }\textbf {\bibinfo {volume} {111}},\ \bibinfo {pages} {080502}
  (\bibinfo {year} {2013})}\BibitemShut {NoStop}%
\bibitem [{\citenamefont {Chen}\ \emph {et~al.}(2014)\citenamefont {Chen} \emph
  {et~al.}}]{chen2014qubit}%
  \BibitemOpen
  \bibfield  {author} {\bibinfo {author} {\bibfnamefont {Y.}~\bibnamefont
  {Chen}} \emph {et~al.},\ }\href@noop {} {\bibfield  {journal} {\bibinfo
  {journal} {Phys. Rev. Lett.}\ }\textbf {\bibinfo {volume} {113}},\ \bibinfo
  {pages} {220502} (\bibinfo {year} {2014})}\BibitemShut {NoStop}%
\bibitem [{\citenamefont {Geller}\ \emph {et~al.}(2014)\citenamefont {Geller},
  \citenamefont {Donate}, \citenamefont {Chen}, \citenamefont {Neill},
  \citenamefont {Roushan},\ and\ \citenamefont {Martinis}}]{geller2014tunable}%
  \BibitemOpen
  \bibfield  {author} {\bibinfo {author} {\bibfnamefont {M.~R.}\ \bibnamefont
  {Geller}}, \bibinfo {author} {\bibfnamefont {E.}~\bibnamefont {Donate}},
  \bibinfo {author} {\bibfnamefont {Y.}~\bibnamefont {Chen}}, \bibinfo {author}
  {\bibfnamefont {C.}~\bibnamefont {Neill}}, \bibinfo {author} {\bibfnamefont
  {P.}~\bibnamefont {Roushan}}, \ and\ \bibinfo {author} {\bibfnamefont
  {J.~M.}\ \bibnamefont {Martinis}},\ }\href@noop {} {\bibfield  {journal}
  {\bibinfo  {journal} {arXiv preprint arXiv:1405.1915}\ } (\bibinfo {year}
  {2014})}\BibitemShut {NoStop}%
\bibitem [{\citenamefont {Neeley}\ \emph {et~al.}(2010)\citenamefont {Neeley},
  \citenamefont {Bialczak}, \citenamefont {Lenander}, \citenamefont {Lucero},
  \citenamefont {Mariantoni}, \citenamefont {O’Connell}, \citenamefont
  {Sank}, \citenamefont {Wang}, \citenamefont {Weides}, \citenamefont {Wenner}
  \emph {et~al.}}]{neeley2010generation}%
  \BibitemOpen
  \bibfield  {author} {\bibinfo {author} {\bibfnamefont {M.}~\bibnamefont
  {Neeley}}, \bibinfo {author} {\bibfnamefont {R.~C.}\ \bibnamefont
  {Bialczak}}, \bibinfo {author} {\bibfnamefont {M.}~\bibnamefont {Lenander}},
  \bibinfo {author} {\bibfnamefont {E.}~\bibnamefont {Lucero}}, \bibinfo
  {author} {\bibfnamefont {M.}~\bibnamefont {Mariantoni}}, \bibinfo {author}
  {\bibfnamefont {A.}~\bibnamefont {O’Connell}}, \bibinfo {author}
  {\bibfnamefont {D.}~\bibnamefont {Sank}}, \bibinfo {author} {\bibfnamefont
  {H.}~\bibnamefont {Wang}}, \bibinfo {author} {\bibfnamefont {M.}~\bibnamefont
  {Weides}}, \bibinfo {author} {\bibfnamefont {J.}~\bibnamefont {Wenner}},
  \emph {et~al.},\ }\href@noop {} {\bibfield  {journal} {\bibinfo  {journal}
  {Nature}\ }\textbf {\bibinfo {volume} {467}},\ \bibinfo {pages} {570}
  (\bibinfo {year} {2010})}\BibitemShut {NoStop}%
\bibitem [{\citenamefont {Paik}\ \emph {et~al.}(2011)\citenamefont {Paik},
  \citenamefont {Schuster}, \citenamefont {Bishop}, \citenamefont {Kirchmair},
  \citenamefont {Catelani}, \citenamefont {Sears}, \citenamefont {Johnson},
  \citenamefont {Reagor}, \citenamefont {Frunzio}, \citenamefont {Glazman}
  \emph {et~al.}}]{paik2011observation}%
  \BibitemOpen
  \bibfield  {author} {\bibinfo {author} {\bibfnamefont {H.}~\bibnamefont
  {Paik}}, \bibinfo {author} {\bibfnamefont {D.}~\bibnamefont {Schuster}},
  \bibinfo {author} {\bibfnamefont {L.~S.}\ \bibnamefont {Bishop}}, \bibinfo
  {author} {\bibfnamefont {G.}~\bibnamefont {Kirchmair}}, \bibinfo {author}
  {\bibfnamefont {G.}~\bibnamefont {Catelani}}, \bibinfo {author}
  {\bibfnamefont {A.}~\bibnamefont {Sears}}, \bibinfo {author} {\bibfnamefont
  {B.}~\bibnamefont {Johnson}}, \bibinfo {author} {\bibfnamefont
  {M.}~\bibnamefont {Reagor}}, \bibinfo {author} {\bibfnamefont
  {L.}~\bibnamefont {Frunzio}}, \bibinfo {author} {\bibfnamefont
  {L.}~\bibnamefont {Glazman}},  \emph {et~al.},\ }\href@noop {} {\bibfield
  {journal} {\bibinfo  {journal} {Physical Review Letters}\ }\textbf {\bibinfo
  {volume} {107}},\ \bibinfo {pages} {240501} (\bibinfo {year}
  {2011})}\BibitemShut {NoStop}%
\bibitem [{\citenamefont {Vedral}\ \emph {et~al.}(1996)\citenamefont {Vedral},
  \citenamefont {Barenco},\ and\ \citenamefont {Ekert}}]{vedral1996quantum}%
  \BibitemOpen
  \bibfield  {author} {\bibinfo {author} {\bibfnamefont {V.}~\bibnamefont
  {Vedral}}, \bibinfo {author} {\bibfnamefont {A.}~\bibnamefont {Barenco}}, \
  and\ \bibinfo {author} {\bibfnamefont {A.}~\bibnamefont {Ekert}},\
  }\href@noop {} {\bibfield  {journal} {\bibinfo  {journal} {Physical Review
  A}\ }\textbf {\bibinfo {volume} {54}},\ \bibinfo {pages} {147} (\bibinfo
  {year} {1996})}\BibitemShut {NoStop}%
\bibitem [{\citenamefont {Fedorov}\ \emph {et~al.}(2011)\citenamefont
  {Fedorov}, \citenamefont {Steffen}, \citenamefont {Baur}, \citenamefont
  {Da~Silva},\ and\ \citenamefont {Wallraff}}]{fedorov2011implementation}%
  \BibitemOpen
  \bibfield  {author} {\bibinfo {author} {\bibfnamefont {A.}~\bibnamefont
  {Fedorov}}, \bibinfo {author} {\bibfnamefont {L.}~\bibnamefont {Steffen}},
  \bibinfo {author} {\bibfnamefont {M.}~\bibnamefont {Baur}}, \bibinfo {author}
  {\bibfnamefont {M.}~\bibnamefont {Da~Silva}}, \ and\ \bibinfo {author}
  {\bibfnamefont {A.}~\bibnamefont {Wallraff}},\ }\href@noop {} {\bibfield
  {journal} {\bibinfo  {journal} {Nature}\ }\textbf {\bibinfo {volume} {481}},\
  \bibinfo {pages} {170} (\bibinfo {year} {2011})}\BibitemShut {NoStop}%
\bibitem [{\citenamefont {Cory}\ \emph {et~al.}(1998)\citenamefont {Cory},
  \citenamefont {Price}, \citenamefont {Maas}, \citenamefont {Knill},
  \citenamefont {Laflamme}, \citenamefont {Zurek}, \citenamefont {Havel},\ and\
  \citenamefont {Somaroo}}]{cory1998experimental}%
  \BibitemOpen
  \bibfield  {author} {\bibinfo {author} {\bibfnamefont {D.~G.}\ \bibnamefont
  {Cory}}, \bibinfo {author} {\bibfnamefont {M.}~\bibnamefont {Price}},
  \bibinfo {author} {\bibfnamefont {W.}~\bibnamefont {Maas}}, \bibinfo {author}
  {\bibfnamefont {E.}~\bibnamefont {Knill}}, \bibinfo {author} {\bibfnamefont
  {R.}~\bibnamefont {Laflamme}}, \bibinfo {author} {\bibfnamefont {W.~H.}\
  \bibnamefont {Zurek}}, \bibinfo {author} {\bibfnamefont {T.~F.}\ \bibnamefont
  {Havel}}, \ and\ \bibinfo {author} {\bibfnamefont {S.}~\bibnamefont
  {Somaroo}},\ }\href@noop {} {\bibfield  {journal} {\bibinfo  {journal}
  {Physical Review Letters}\ }\textbf {\bibinfo {volume} {81}},\ \bibinfo
  {pages} {2152} (\bibinfo {year} {1998})}\BibitemShut {NoStop}%
\bibitem [{\citenamefont {Bravyi}\ \emph {et~al.}(2008)\citenamefont {Bravyi},
  \citenamefont {DiVincenzo}, \citenamefont {Loss},\ and\ \citenamefont
  {Terhal}}]{bravyi2008quantum}%
  \BibitemOpen
  \bibfield  {author} {\bibinfo {author} {\bibfnamefont {S.}~\bibnamefont
  {Bravyi}}, \bibinfo {author} {\bibfnamefont {D.~P.}\ \bibnamefont
  {DiVincenzo}}, \bibinfo {author} {\bibfnamefont {D.}~\bibnamefont {Loss}}, \
  and\ \bibinfo {author} {\bibfnamefont {B.~M.}\ \bibnamefont {Terhal}},\
  }\href@noop {} {\bibfield  {journal} {\bibinfo  {journal} {Phys. Rev. Lett.}\
  }\textbf {\bibinfo {volume} {101}},\ \bibinfo {pages} {070503} (\bibinfo
  {year} {2008})}\BibitemShut {NoStop}%
\bibitem [{\citenamefont {Biamonte}\ and\ \citenamefont
  {Love}(2008)}]{biamonte2008realizable}%
  \BibitemOpen
  \bibfield  {author} {\bibinfo {author} {\bibfnamefont {J.~D.}\ \bibnamefont
  {Biamonte}}\ and\ \bibinfo {author} {\bibfnamefont {P.~J.}\ \bibnamefont
  {Love}},\ }\href@noop {} {\bibfield  {journal} {\bibinfo  {journal} {Physical
  Review A}\ }\textbf {\bibinfo {volume} {78}},\ \bibinfo {pages} {012352}
  (\bibinfo {year} {2008})}\BibitemShut {NoStop}%
\bibitem [{\citenamefont {Brif}\ \emph {et~al.}(2010)\citenamefont {Brif},
  \citenamefont {Chakrabarti},\ and\ \citenamefont {Rabitz}}]{brif2010control}%
  \BibitemOpen
  \bibfield  {author} {\bibinfo {author} {\bibfnamefont {C.}~\bibnamefont
  {Brif}}, \bibinfo {author} {\bibfnamefont {R.}~\bibnamefont {Chakrabarti}}, \
  and\ \bibinfo {author} {\bibfnamefont {H.}~\bibnamefont {Rabitz}},\
  }\href@noop {} {\bibfield  {journal} {\bibinfo  {journal} {New Journal of
  Physics}\ }\textbf {\bibinfo {volume} {12}},\ \bibinfo {pages} {075008}
  (\bibinfo {year} {2010})}\BibitemShut {NoStop}%
\bibitem [{\citenamefont {Bishop}(2006)}]{bishop2006pattern}%
  \BibitemOpen
  \bibfield  {author} {\bibinfo {author} {\bibfnamefont {C.~M.}\ \bibnamefont
  {Bishop}},\ }\href@noop {} {\emph {\bibinfo {title} {Pattern recognition and
  machine learning}}}\ (\bibinfo  {publisher} {Springer},\ \bibinfo {year}
  {2006})\BibitemShut {NoStop}%
\bibitem [{\citenamefont {Wittek}(2014)}]{wittek2014quantum}%
  \BibitemOpen
  \bibfield  {author} {\bibinfo {author} {\bibfnamefont {P.}~\bibnamefont
  {Wittek}},\ }\href@noop {} {\emph {\bibinfo {title} {Quantum machine
  learning: what quantum computing means to data mining}}}\ (\bibinfo
  {publisher} {Elsevier, Oxford},\ \bibinfo {year} {2014})\BibitemShut
  {NoStop}%
\bibitem [{\citenamefont {Rebentrost}\ \emph {et~al.}(2014)\citenamefont
  {Rebentrost}, \citenamefont {Mohseni},\ and\ \citenamefont
  {Lloyd}}]{rebentrost2014quantum}%
  \BibitemOpen
  \bibfield  {author} {\bibinfo {author} {\bibfnamefont {P.}~\bibnamefont
  {Rebentrost}}, \bibinfo {author} {\bibfnamefont {M.}~\bibnamefont {Mohseni}},
  \ and\ \bibinfo {author} {\bibfnamefont {S.}~\bibnamefont {Lloyd}},\
  }\href@noop {} {\bibfield  {journal} {\bibinfo  {journal} {Phys. Rev. Lett.}\
  }\textbf {\bibinfo {volume} {113}},\ \bibinfo {pages} {130503} (\bibinfo
  {year} {2014})}\BibitemShut {NoStop}%
\bibitem [{\citenamefont {Paparo}\ \emph {et~al.}(2014)\citenamefont {Paparo},
  \citenamefont {Dunjko}, \citenamefont {Makmal}, \citenamefont
  {Martin-Delgado},\ and\ \citenamefont {Briegel}}]{paparo2014quantum}%
  \BibitemOpen
  \bibfield  {author} {\bibinfo {author} {\bibfnamefont {G.~D.}\ \bibnamefont
  {Paparo}}, \bibinfo {author} {\bibfnamefont {V.}~\bibnamefont {Dunjko}},
  \bibinfo {author} {\bibfnamefont {A.}~\bibnamefont {Makmal}}, \bibinfo
  {author} {\bibfnamefont {M.~A.}\ \bibnamefont {Martin-Delgado}}, \ and\
  \bibinfo {author} {\bibfnamefont {H.~J.}\ \bibnamefont {Briegel}},\
  }\href@noop {} {\bibfield  {journal} {\bibinfo  {journal} {Physical Review
  X}\ }\textbf {\bibinfo {volume} {4}},\ \bibinfo {pages} {031002} (\bibinfo
  {year} {2014})}\BibitemShut {NoStop}%
\bibitem [{\citenamefont {Wiebe}\ \emph {et~al.}(2015)\citenamefont {Wiebe},
  \citenamefont {Kapoor},\ and\ \citenamefont {Svore}}]{wiebe2014quantum}%
  \BibitemOpen
  \bibfield  {author} {\bibinfo {author} {\bibfnamefont {N.}~\bibnamefont
  {Wiebe}}, \bibinfo {author} {\bibfnamefont {A.}~\bibnamefont {Kapoor}}, \
  and\ \bibinfo {author} {\bibfnamefont {K.}~\bibnamefont {Svore}},\
  }\href@noop {} {\bibfield  {journal} {\bibinfo  {journal} {Quantum
  Information \& Computation}\ }\textbf {\bibinfo {volume} {15}},\ \bibinfo
  {pages} {0318} (\bibinfo {year} {2015})}\BibitemShut {NoStop}%
\bibitem [{\citenamefont {Lloyd}\ \emph {et~al.}(2014)\citenamefont {Lloyd},
  \citenamefont {Mohseni},\ and\ \citenamefont
  {Rebentrost}}]{lloyd2014quantum}%
  \BibitemOpen
  \bibfield  {author} {\bibinfo {author} {\bibfnamefont {S.}~\bibnamefont
  {Lloyd}}, \bibinfo {author} {\bibfnamefont {M.}~\bibnamefont {Mohseni}}, \
  and\ \bibinfo {author} {\bibfnamefont {P.}~\bibnamefont {Rebentrost}},\
  }\href@noop {} {\bibfield  {journal} {\bibinfo  {journal} {Nature Physics}\
  }\textbf {\bibinfo {volume} {10}},\ \bibinfo {pages} {631} (\bibinfo {year}
  {2014})}\BibitemShut {NoStop}%
\bibitem [{\citenamefont {Rotondo}\ \emph {et~al.}(2015)\citenamefont
  {Rotondo}, \citenamefont {Lagomarsino},\ and\ \citenamefont
  {Viola}}]{rotondo2015dicke}%
  \BibitemOpen
  \bibfield  {author} {\bibinfo {author} {\bibfnamefont {P.}~\bibnamefont
  {Rotondo}}, \bibinfo {author} {\bibfnamefont {M.~C.}\ \bibnamefont
  {Lagomarsino}}, \ and\ \bibinfo {author} {\bibfnamefont {G.}~\bibnamefont
  {Viola}},\ }\href@noop {} {\bibfield  {journal} {\bibinfo  {journal}
  {Physical review letters}\ }\textbf {\bibinfo {volume} {114}},\ \bibinfo
  {pages} {143601} (\bibinfo {year} {2015})}\BibitemShut {NoStop}%
\bibitem [{\citenamefont {Nagaj}(2012)}]{nagaj2012universal}%
  \BibitemOpen
  \bibfield  {author} {\bibinfo {author} {\bibfnamefont {D.}~\bibnamefont
  {Nagaj}},\ }\href@noop {} {\bibfield  {journal} {\bibinfo  {journal}
  {Physical Review A}\ }\textbf {\bibinfo {volume} {85}},\ \bibinfo {pages}
  {032330} (\bibinfo {year} {2012})}\BibitemShut {NoStop}%
\bibitem [{\citenamefont {Bang}\ \emph {et~al.}(2008)\citenamefont {Bang},
  \citenamefont {Lim}, \citenamefont {Kim},\ and\ \citenamefont
  {Lee}}]{bang2008quantum}%
  \BibitemOpen
  \bibfield  {author} {\bibinfo {author} {\bibfnamefont {J.}~\bibnamefont
  {Bang}}, \bibinfo {author} {\bibfnamefont {J.}~\bibnamefont {Lim}}, \bibinfo
  {author} {\bibfnamefont {M.}~\bibnamefont {Kim}}, \ and\ \bibinfo {author}
  {\bibfnamefont {J.}~\bibnamefont {Lee}},\ }\href@noop {} {\bibfield
  {journal} {\bibinfo  {journal} {arXiv preprint arXiv:0803.2976}\ } (\bibinfo
  {year} {2008})}\BibitemShut {NoStop}%
\bibitem [{\citenamefont {Gammelmark}\ and\ \citenamefont
  {M{\o}lmer}(2009)}]{gammelmark2009quantum}%
  \BibitemOpen
  \bibfield  {author} {\bibinfo {author} {\bibfnamefont {S.}~\bibnamefont
  {Gammelmark}}\ and\ \bibinfo {author} {\bibfnamefont {K.}~\bibnamefont
  {M{\o}lmer}},\ }\href@noop {} {\bibfield  {journal} {\bibinfo  {journal} {New
  Journal of Physics}\ }\textbf {\bibinfo {volume} {11}},\ \bibinfo {pages}
  {033017} (\bibinfo {year} {2009})}\BibitemShut {NoStop}%
\bibitem [{\citenamefont {Magesan}\ \emph {et~al.}(2011)\citenamefont
  {Magesan}, \citenamefont {Blume-Kohout},\ and\ \citenamefont
  {Emerson}}]{magesan2011gate}%
  \BibitemOpen
  \bibfield  {author} {\bibinfo {author} {\bibfnamefont {E.}~\bibnamefont
  {Magesan}}, \bibinfo {author} {\bibfnamefont {R.}~\bibnamefont
  {Blume-Kohout}}, \ and\ \bibinfo {author} {\bibfnamefont {J.}~\bibnamefont
  {Emerson}},\ }\href@noop {} {\bibfield  {journal} {\bibinfo  {journal}
  {Physical Review A}\ }\textbf {\bibinfo {volume} {84}},\ \bibinfo {pages}
  {012309} (\bibinfo {year} {2011})}\BibitemShut {NoStop}%
\bibitem [{\citenamefont {Pedersen}\ \emph {et~al.}(2008)\citenamefont
  {Pedersen}, \citenamefont {M{\o}ller},\ and\ \citenamefont
  {M{\o}lmer}}]{pedersen2008distribution}%
  \BibitemOpen
  \bibfield  {author} {\bibinfo {author} {\bibfnamefont {L.~H.}\ \bibnamefont
  {Pedersen}}, \bibinfo {author} {\bibfnamefont {N.~M.}\ \bibnamefont
  {M{\o}ller}}, \ and\ \bibinfo {author} {\bibfnamefont {K.}~\bibnamefont
  {M{\o}lmer}},\ }\href@noop {} {\bibfield  {journal} {\bibinfo  {journal}
  {Physics Letters A}\ }\textbf {\bibinfo {volume} {372}},\ \bibinfo {pages}
  {7028} (\bibinfo {year} {2008})}\BibitemShut {NoStop}%
\bibitem [{\citenamefont {Collins}\ and\ \citenamefont
  {{\'S}niady}(2006)}]{collins2006integration}%
  \BibitemOpen
  \bibfield  {author} {\bibinfo {author} {\bibfnamefont {B.}~\bibnamefont
  {Collins}}\ and\ \bibinfo {author} {\bibfnamefont {P.}~\bibnamefont
  {{\'S}niady}},\ }\href@noop {} {\bibfield  {journal} {\bibinfo  {journal}
  {Communications in Mathematical Physics}\ }\textbf {\bibinfo {volume}
  {264}},\ \bibinfo {pages} {773} (\bibinfo {year} {2006})}\BibitemShut
  {NoStop}%
\bibitem [{\citenamefont {Bottou}(1998)}]{bottou1998online}%
  \BibitemOpen
  \bibfield  {author} {\bibinfo {author} {\bibfnamefont {L.}~\bibnamefont
  {Bottou}},\ }in\ \href@noop {} {\emph {\bibinfo {booktitle} {Online Learning
  and Neural Networks}}},\ \bibinfo {editor} {edited by\ \bibinfo {editor}
  {\bibfnamefont {D.}~\bibnamefont {Saad}}}\ (\bibinfo  {publisher} {Cambridge
  University Press},\ \bibinfo {address} {Cambridge, UK},\ \bibinfo {year}
  {1998})\ \bibinfo {note} {revised, oct 2012}\BibitemShut {NoStop}%
\bibitem [{\citenamefont {Johansson}\ \emph {et~al.}(2013)\citenamefont
  {Johansson}, \citenamefont {Nation},\ and\ \citenamefont
  {Nori}}]{Johansson20131234}%
  \BibitemOpen
  \bibfield  {author} {\bibinfo {author} {\bibfnamefont {J.}~\bibnamefont
  {Johansson}}, \bibinfo {author} {\bibfnamefont {P.}~\bibnamefont {Nation}}, \
  and\ \bibinfo {author} {\bibfnamefont {F.}~\bibnamefont {Nori}},\ }\href
  {\doibase 10.1016/j.cpc.2012.11.019} {\bibfield  {journal} {\bibinfo
  {journal} {Computer Physics Communications}\ }\textbf {\bibinfo {volume}
  {184}},\ \bibinfo {pages} {1234 } (\bibinfo {year} {2013})}\BibitemShut
  {NoStop}%
\bibitem [{\citenamefont {Shor}(1997)}]{shor1997polynomial}%
  \BibitemOpen
  \bibfield  {author} {\bibinfo {author} {\bibfnamefont {P.~W.}\ \bibnamefont
  {Shor}},\ }\href@noop {} {\bibfield  {journal} {\bibinfo  {journal} {SIAM
  journal on computing}\ }\textbf {\bibinfo {volume} {26}},\ \bibinfo {pages}
  {1484} (\bibinfo {year} {1997})}\BibitemShut {NoStop}%
\bibitem [{\citenamefont {Dennis}(2001)}]{dennis2001toward}%
  \BibitemOpen
  \bibfield  {author} {\bibinfo {author} {\bibfnamefont {E.}~\bibnamefont
  {Dennis}},\ }\href@noop {} {\bibfield  {journal} {\bibinfo  {journal}
  {Physical Review A}\ }\textbf {\bibinfo {volume} {63}},\ \bibinfo {pages}
  {052314} (\bibinfo {year} {2001})}\BibitemShut {NoStop}%
\bibitem [{\citenamefont {Shi}(2003)}]{shi2003both}%
  \BibitemOpen
  \bibfield  {author} {\bibinfo {author} {\bibfnamefont {Y.}~\bibnamefont
  {Shi}},\ }\href@noop {} {\bibfield  {journal} {\bibinfo  {journal} {Quantum
  Information \& Computation}\ }\textbf {\bibinfo {volume} {3}},\ \bibinfo
  {pages} {84} (\bibinfo {year} {2003})}\BibitemShut {NoStop}%
\bibitem [{\citenamefont {Monz}\ \emph {et~al.}(2009)\citenamefont {Monz},
  \citenamefont {Kim}, \citenamefont {H{\"a}nsel}, \citenamefont {Riebe},
  \citenamefont {Villar}, \citenamefont {Schindler}, \citenamefont {Chwalla},
  \citenamefont {Hennrich},\ and\ \citenamefont {Blatt}}]{monz2009realization}%
  \BibitemOpen
  \bibfield  {author} {\bibinfo {author} {\bibfnamefont {T.}~\bibnamefont
  {Monz}}, \bibinfo {author} {\bibfnamefont {K.}~\bibnamefont {Kim}}, \bibinfo
  {author} {\bibfnamefont {W.}~\bibnamefont {H{\"a}nsel}}, \bibinfo {author}
  {\bibfnamefont {M.}~\bibnamefont {Riebe}}, \bibinfo {author} {\bibfnamefont
  {A.}~\bibnamefont {Villar}}, \bibinfo {author} {\bibfnamefont
  {P.}~\bibnamefont {Schindler}}, \bibinfo {author} {\bibfnamefont
  {M.}~\bibnamefont {Chwalla}}, \bibinfo {author} {\bibfnamefont
  {M.}~\bibnamefont {Hennrich}}, \ and\ \bibinfo {author} {\bibfnamefont
  {R.}~\bibnamefont {Blatt}},\ }\href@noop {} {\bibfield  {journal} {\bibinfo
  {journal} {Physical review letters}\ }\textbf {\bibinfo {volume} {102}},\
  \bibinfo {pages} {040501} (\bibinfo {year} {2009})}\BibitemShut {NoStop}%
\bibitem [{\citenamefont {Reed}\ \emph {et~al.}(2012)\citenamefont {Reed},
  \citenamefont {DiCarlo}, \citenamefont {Nigg}, \citenamefont {Sun},
  \citenamefont {Frunzio}, \citenamefont {Girvin},\ and\ \citenamefont
  {Schoelkopf}}]{reed2012realization}%
  \BibitemOpen
  \bibfield  {author} {\bibinfo {author} {\bibfnamefont {M.}~\bibnamefont
  {Reed}}, \bibinfo {author} {\bibfnamefont {L.}~\bibnamefont {DiCarlo}},
  \bibinfo {author} {\bibfnamefont {S.}~\bibnamefont {Nigg}}, \bibinfo {author}
  {\bibfnamefont {L.}~\bibnamefont {Sun}}, \bibinfo {author} {\bibfnamefont
  {L.}~\bibnamefont {Frunzio}}, \bibinfo {author} {\bibfnamefont
  {S.}~\bibnamefont {Girvin}}, \ and\ \bibinfo {author} {\bibfnamefont
  {R.}~\bibnamefont {Schoelkopf}},\ }\href@noop {} {\bibfield  {journal}
  {\bibinfo  {journal} {Nature}\ }\textbf {\bibinfo {volume} {482}},\ \bibinfo
  {pages} {382} (\bibinfo {year} {2012})}\BibitemShut {NoStop}%
\bibitem [{\citenamefont {Lanyon}\ \emph {et~al.}(2009)\citenamefont {Lanyon},
  \citenamefont {Barbieri}, \citenamefont {Almeida}, \citenamefont {Jennewein},
  \citenamefont {Ralph}, \citenamefont {Resch}, \citenamefont {Pryde},
  \citenamefont {O'Brien}, \citenamefont {Gilchrist},\ and\ \citenamefont
  {White}}]{lanyon2009simplifying}%
  \BibitemOpen
  \bibfield  {author} {\bibinfo {author} {\bibfnamefont {B.~P.}\ \bibnamefont
  {Lanyon}}, \bibinfo {author} {\bibfnamefont {M.}~\bibnamefont {Barbieri}},
  \bibinfo {author} {\bibfnamefont {M.~P.}\ \bibnamefont {Almeida}}, \bibinfo
  {author} {\bibfnamefont {T.}~\bibnamefont {Jennewein}}, \bibinfo {author}
  {\bibfnamefont {T.~C.}\ \bibnamefont {Ralph}}, \bibinfo {author}
  {\bibfnamefont {K.~J.}\ \bibnamefont {Resch}}, \bibinfo {author}
  {\bibfnamefont {G.~J.}\ \bibnamefont {Pryde}}, \bibinfo {author}
  {\bibfnamefont {J.~L.}\ \bibnamefont {O'Brien}}, \bibinfo {author}
  {\bibfnamefont {A.}~\bibnamefont {Gilchrist}}, \ and\ \bibinfo {author}
  {\bibfnamefont {A.~G.}\ \bibnamefont {White}},\ }\href@noop {} {\bibfield
  {journal} {\bibinfo  {journal} {Nature Physics}\ }\textbf {\bibinfo {volume}
  {5}},\ \bibinfo {pages} {134} (\bibinfo {year} {2009})}\BibitemShut {NoStop}%
\bibitem [{\citenamefont {Stojanovi{\'c}}\ \emph {et~al.}(2012)\citenamefont
  {Stojanovi{\'c}}, \citenamefont {Fedorov}, \citenamefont {Wallraff},\ and\
  \citenamefont {Bruder}}]{stojanovic2012quantum}%
  \BibitemOpen
  \bibfield  {author} {\bibinfo {author} {\bibfnamefont {V.~M.}\ \bibnamefont
  {Stojanovi{\'c}}}, \bibinfo {author} {\bibfnamefont {A.}~\bibnamefont
  {Fedorov}}, \bibinfo {author} {\bibfnamefont {A.}~\bibnamefont {Wallraff}}, \
  and\ \bibinfo {author} {\bibfnamefont {C.}~\bibnamefont {Bruder}},\
  }\href@noop {} {\bibfield  {journal} {\bibinfo  {journal} {Physical Review
  B}\ }\textbf {\bibinfo {volume} {85}},\ \bibinfo {pages} {054504} (\bibinfo
  {year} {2012})}\BibitemShut {NoStop}%
\bibitem [{\citenamefont {Zahedinejad}\ \emph {et~al.}(2015)\citenamefont
  {Zahedinejad}, \citenamefont {Ghosh},\ and\ \citenamefont
  {Sanders}}]{zahedinejad2015high}%
  \BibitemOpen
  \bibfield  {author} {\bibinfo {author} {\bibfnamefont {E.}~\bibnamefont
  {Zahedinejad}}, \bibinfo {author} {\bibfnamefont {J.}~\bibnamefont {Ghosh}},
  \ and\ \bibinfo {author} {\bibfnamefont {B.~C.}\ \bibnamefont {Sanders}},\
  }\href@noop {} {\bibfield  {journal} {\bibinfo  {journal} {Physical Review
  Letters}\ }\textbf {\bibinfo {volume} {114}},\ \bibinfo {pages} {200502}
  (\bibinfo {year} {2015})}\BibitemShut {NoStop}%
\bibitem [{\citenamefont {Gaitan}(2008)}]{gaitan2008quantum}%
  \BibitemOpen
  \bibfield  {author} {\bibinfo {author} {\bibfnamefont {F.}~\bibnamefont
  {Gaitan}},\ }\href@noop {} {\emph {\bibinfo {title} {Quantum error correction
  and fault tolerant quantum computing}}}\ (\bibinfo  {publisher} {CRC Press},\
  \bibinfo {year} {2008})\BibitemShut {NoStop}%
\bibitem [{\citenamefont {Yao}\ \emph {et~al.}(2013)\citenamefont {Yao},
  \citenamefont {Gong}, \citenamefont {Laumann}, \citenamefont {Bennett},
  \citenamefont {Duan}, \citenamefont {Lukin}, \citenamefont {Jiang},\ and\
  \citenamefont {Gorshkov}}]{yao2013quantum}%
  \BibitemOpen
  \bibfield  {author} {\bibinfo {author} {\bibfnamefont {N.~Y.}\ \bibnamefont
  {Yao}}, \bibinfo {author} {\bibfnamefont {Z.-X.}\ \bibnamefont {Gong}},
  \bibinfo {author} {\bibfnamefont {C.~R.}\ \bibnamefont {Laumann}}, \bibinfo
  {author} {\bibfnamefont {S.~D.}\ \bibnamefont {Bennett}}, \bibinfo {author}
  {\bibfnamefont {L.-M.}\ \bibnamefont {Duan}}, \bibinfo {author}
  {\bibfnamefont {M.~D.}\ \bibnamefont {Lukin}}, \bibinfo {author}
  {\bibfnamefont {L.}~\bibnamefont {Jiang}}, \ and\ \bibinfo {author}
  {\bibfnamefont {A.~V.}\ \bibnamefont {Gorshkov}},\ }\href@noop {} {\bibfield
  {journal} {\bibinfo  {journal} {Physical Review A}\ }\textbf {\bibinfo
  {volume} {87}},\ \bibinfo {pages} {022306} (\bibinfo {year}
  {2013})}\BibitemShut {NoStop}%
\bibitem [{\citenamefont {Banchi}\ \emph {et~al.}(2015)\citenamefont {Banchi},
  \citenamefont {Compagno},\ and\ \citenamefont {Bose}}]{banchi2015perfect}%
  \BibitemOpen
  \bibfield  {author} {\bibinfo {author} {\bibfnamefont {L.}~\bibnamefont
  {Banchi}}, \bibinfo {author} {\bibfnamefont {E.}~\bibnamefont {Compagno}}, \
  and\ \bibinfo {author} {\bibfnamefont {S.}~\bibnamefont {Bose}},\ }\href@noop
  {} {\bibfield  {journal} {\bibinfo  {journal} {Physical Review A}\ }\textbf
  {\bibinfo {volume} {91}},\ \bibinfo {pages} {052323} (\bibinfo {year}
  {2015})}\BibitemShut {NoStop}%
\bibitem [{\citenamefont {Yao}\ \emph {et~al.}(2012)\citenamefont {Yao},
  \citenamefont {Jiang}, \citenamefont {Gorshkov}, \citenamefont {Maurer},
  \citenamefont {Giedke}, \citenamefont {Cirac},\ and\ \citenamefont
  {Lukin}}]{yao2012scalable}%
  \BibitemOpen
  \bibfield  {author} {\bibinfo {author} {\bibfnamefont {N.~Y.}\ \bibnamefont
  {Yao}}, \bibinfo {author} {\bibfnamefont {L.}~\bibnamefont {Jiang}}, \bibinfo
  {author} {\bibfnamefont {A.~V.}\ \bibnamefont {Gorshkov}}, \bibinfo {author}
  {\bibfnamefont {P.~C.}\ \bibnamefont {Maurer}}, \bibinfo {author}
  {\bibfnamefont {G.}~\bibnamefont {Giedke}}, \bibinfo {author} {\bibfnamefont
  {J.~I.}\ \bibnamefont {Cirac}}, \ and\ \bibinfo {author} {\bibfnamefont
  {M.~D.}\ \bibnamefont {Lukin}},\ }\href@noop {} {\bibfield  {journal}
  {\bibinfo  {journal} {Nature communications}\ }\textbf {\bibinfo {volume}
  {3}},\ \bibinfo {pages} {800} (\bibinfo {year} {2012})}\BibitemShut {NoStop}%
\bibitem [{\citenamefont {Barends}\ \emph {et~al.}(2014)\citenamefont
  {Barends}, \citenamefont {Kelly}, \citenamefont {Megrant}, \citenamefont
  {Veitia}, \citenamefont {Sank}, \citenamefont {Jeffrey}, \citenamefont
  {White}, \citenamefont {Mutus}, \citenamefont {Fowler}, \citenamefont
  {Campbell} \emph {et~al.}}]{barends2014superconducting}%
  \BibitemOpen
  \bibfield  {author} {\bibinfo {author} {\bibfnamefont {R.}~\bibnamefont
  {Barends}}, \bibinfo {author} {\bibfnamefont {J.}~\bibnamefont {Kelly}},
  \bibinfo {author} {\bibfnamefont {A.}~\bibnamefont {Megrant}}, \bibinfo
  {author} {\bibfnamefont {A.}~\bibnamefont {Veitia}}, \bibinfo {author}
  {\bibfnamefont {D.}~\bibnamefont {Sank}}, \bibinfo {author} {\bibfnamefont
  {E.}~\bibnamefont {Jeffrey}}, \bibinfo {author} {\bibfnamefont
  {T.}~\bibnamefont {White}}, \bibinfo {author} {\bibfnamefont
  {J.}~\bibnamefont {Mutus}}, \bibinfo {author} {\bibfnamefont
  {A.}~\bibnamefont {Fowler}}, \bibinfo {author} {\bibfnamefont
  {B.}~\bibnamefont {Campbell}},  \emph {et~al.},\ }\href@noop {} {\bibfield
  {journal} {\bibinfo  {journal} {Nature}\ }\textbf {\bibinfo {volume} {508}},\
  \bibinfo {pages} {500} (\bibinfo {year} {2014})}\BibitemShut {NoStop}%
\bibitem [{\citenamefont {Monz}\ \emph {et~al.}(2016)\citenamefont {Monz},
  \citenamefont {Nigg}, \citenamefont {Martinez}, \citenamefont {Brandl},
  \citenamefont {Schindler}, \citenamefont {Rines}, \citenamefont {Wang},
  \citenamefont {Chuang},\ and\ \citenamefont {Blatt}}]{Monz1068}%
  \BibitemOpen
  \bibfield  {author} {\bibinfo {author} {\bibfnamefont {T.}~\bibnamefont
  {Monz}}, \bibinfo {author} {\bibfnamefont {D.}~\bibnamefont {Nigg}}, \bibinfo
  {author} {\bibfnamefont {E.~A.}\ \bibnamefont {Martinez}}, \bibinfo {author}
  {\bibfnamefont {M.~F.}\ \bibnamefont {Brandl}}, \bibinfo {author}
  {\bibfnamefont {P.}~\bibnamefont {Schindler}}, \bibinfo {author}
  {\bibfnamefont {R.}~\bibnamefont {Rines}}, \bibinfo {author} {\bibfnamefont
  {S.~X.}\ \bibnamefont {Wang}}, \bibinfo {author} {\bibfnamefont {I.~L.}\
  \bibnamefont {Chuang}}, \ and\ \bibinfo {author} {\bibfnamefont
  {R.}~\bibnamefont {Blatt}},\ }\href {\doibase 10.1126/science.aad9480}
  {\bibfield  {journal} {\bibinfo  {journal} {Science}\ }\textbf {\bibinfo
  {volume} {351}},\ \bibinfo {pages} {1068} (\bibinfo {year}
  {2016})}\BibitemShut {NoStop}%
\bibitem [{\citenamefont {Mariantoni}\ \emph {et~al.}(2011)\citenamefont
  {Mariantoni} \emph {et~al.}}]{mariantoni2011implementing}%
  \BibitemOpen
  \bibfield  {author} {\bibinfo {author} {\bibfnamefont {M.}~\bibnamefont
  {Mariantoni}} \emph {et~al.},\ }\href@noop {} {\bibfield  {journal} {\bibinfo
   {journal} {Science}\ }\textbf {\bibinfo {volume} {334}},\ \bibinfo {pages}
  {61} (\bibinfo {year} {2011})}\BibitemShut {NoStop}%
\bibitem [{\citenamefont {Nikolopoulos}\ and\ \citenamefont
  {Jex}(2014)}]{nikolopoulos2014quantum}%
  \BibitemOpen
  \bibfield  {author} {\bibinfo {author} {\bibfnamefont {G.~M.}\ \bibnamefont
  {Nikolopoulos}}\ and\ \bibinfo {author} {\bibfnamefont {I.}~\bibnamefont
  {Jex}},\ }\href@noop {} {\emph {\bibinfo {title} {Quantum State Transfer and
  Network Engineering}}}\ (\bibinfo  {publisher} {Springer},\ \bibinfo {year}
  {2014})\BibitemShut {NoStop}%
\bibitem [{\citenamefont {Storn}\ and\ \citenamefont
  {Price}(1997)}]{storn1997differential}%
  \BibitemOpen
  \bibfield  {author} {\bibinfo {author} {\bibfnamefont {R.}~\bibnamefont
  {Storn}}\ and\ \bibinfo {author} {\bibfnamefont {K.}~\bibnamefont {Price}},\
  }\href@noop {} {\bibfield  {journal} {\bibinfo  {journal} {Journal of global
  optimization}\ }\textbf {\bibinfo {volume} {11}},\ \bibinfo {pages} {341}
  (\bibinfo {year} {1997})}\BibitemShut {NoStop}%
\bibitem [{\citenamefont {Krizhevsky}\ \emph {et~al.}(2012)\citenamefont
  {Krizhevsky}, \citenamefont {Sutskever},\ and\ \citenamefont
  {Hinton}}]{krizhevsky2012imagenet}%
  \BibitemOpen
  \bibfield  {author} {\bibinfo {author} {\bibfnamefont {A.}~\bibnamefont
  {Krizhevsky}}, \bibinfo {author} {\bibfnamefont {I.}~\bibnamefont
  {Sutskever}}, \ and\ \bibinfo {author} {\bibfnamefont {G.~E.}\ \bibnamefont
  {Hinton}},\ }in\ \href@noop {} {\emph {\bibinfo {booktitle} {Advances in
  Neural Information Processing Systems 25}}},\ \bibinfo {editor} {edited by\
  \bibinfo {editor} {\bibfnamefont {F.}~\bibnamefont {Pereira}}, \bibinfo
  {editor} {\bibfnamefont {C.~J.~C.}\ \bibnamefont {Burges}}, \bibinfo {editor}
  {\bibfnamefont {L.}~\bibnamefont {Bottou}}, \ and\ \bibinfo {editor}
  {\bibfnamefont {K.~Q.}\ \bibnamefont {Weinberger}}}\ (\bibinfo  {publisher}
  {Curran Associates, Inc.},\ \bibinfo {year} {2012})\ pp.\ \bibinfo {pages}
  {1097--1105}\BibitemShut {NoStop}%
\bibitem [{\citenamefont {Bisio}\ \emph {et~al.}(2010)\citenamefont {Bisio},
  \citenamefont {Chiribella}, \citenamefont {D’Ariano}, \citenamefont
  {Facchini},\ and\ \citenamefont {Perinotti}}]{bisio2010optimal}%
  \BibitemOpen
  \bibfield  {author} {\bibinfo {author} {\bibfnamefont {A.}~\bibnamefont
  {Bisio}}, \bibinfo {author} {\bibfnamefont {G.}~\bibnamefont {Chiribella}},
  \bibinfo {author} {\bibfnamefont {G.~M.}\ \bibnamefont {D’Ariano}},
  \bibinfo {author} {\bibfnamefont {S.}~\bibnamefont {Facchini}}, \ and\
  \bibinfo {author} {\bibfnamefont {P.}~\bibnamefont {Perinotti}},\ }\href@noop
  {} {\bibfield  {journal} {\bibinfo  {journal} {Physical Review A}\ }\textbf
  {\bibinfo {volume} {81}},\ \bibinfo {pages} {032324} (\bibinfo {year}
  {2010})}\BibitemShut {NoStop}%
\bibitem [{\citenamefont {Wiebe}\ \emph {et~al.}(2014)\citenamefont {Wiebe},
  \citenamefont {Granade}, \citenamefont {Ferrie},\ and\ \citenamefont
  {Cory}}]{wiebe2014hamiltonian}%
  \BibitemOpen
  \bibfield  {author} {\bibinfo {author} {\bibfnamefont {N.}~\bibnamefont
  {Wiebe}}, \bibinfo {author} {\bibfnamefont {C.}~\bibnamefont {Granade}},
  \bibinfo {author} {\bibfnamefont {C.}~\bibnamefont {Ferrie}}, \ and\ \bibinfo
  {author} {\bibfnamefont {D.}~\bibnamefont {Cory}},\ }\href@noop {} {\bibfield
   {journal} {\bibinfo  {journal} {Physical review letters}\ }\textbf {\bibinfo
  {volume} {112}},\ \bibinfo {pages} {190501} (\bibinfo {year}
  {2014})}\BibitemShut {NoStop}%
\bibitem [{\citenamefont {Lu}\ \emph {et~al.}(2015)\citenamefont {Lu},
  \citenamefont {Li}, \citenamefont {Trottier}, \citenamefont {Li},
  \citenamefont {Brodutch}, \citenamefont {Krismanich}, \citenamefont
  {Ghavami}, \citenamefont {Dmitrienko}, \citenamefont {Long}, \citenamefont
  {Baugh} \emph {et~al.}}]{lu2015experimental}%
  \BibitemOpen
  \bibfield  {author} {\bibinfo {author} {\bibfnamefont {D.}~\bibnamefont
  {Lu}}, \bibinfo {author} {\bibfnamefont {H.}~\bibnamefont {Li}}, \bibinfo
  {author} {\bibfnamefont {D.-A.}\ \bibnamefont {Trottier}}, \bibinfo {author}
  {\bibfnamefont {J.}~\bibnamefont {Li}}, \bibinfo {author} {\bibfnamefont
  {A.}~\bibnamefont {Brodutch}}, \bibinfo {author} {\bibfnamefont {A.~P.}\
  \bibnamefont {Krismanich}}, \bibinfo {author} {\bibfnamefont
  {A.}~\bibnamefont {Ghavami}}, \bibinfo {author} {\bibfnamefont {G.~I.}\
  \bibnamefont {Dmitrienko}}, \bibinfo {author} {\bibfnamefont
  {G.}~\bibnamefont {Long}}, \bibinfo {author} {\bibfnamefont {J.}~\bibnamefont
  {Baugh}},  \emph {et~al.},\ }\href@noop {} {\bibfield  {journal} {\bibinfo
  {journal} {Physical review letters}\ }\textbf {\bibinfo {volume} {114}},\
  \bibinfo {pages} {140505} (\bibinfo {year} {2015})}\BibitemShut {NoStop}%
\bibitem [{\citenamefont {Spall}(2005)}]{spall2005introduction}%
  \BibitemOpen
  \bibfield  {author} {\bibinfo {author} {\bibfnamefont {J.~C.}\ \bibnamefont
  {Spall}},\ }\href@noop {} {\emph {\bibinfo {title} {Introduction to
  stochastic search and optimization: estimation, simulation, and control}}},\
  Vol.~\bibinfo {volume} {65}\ (\bibinfo  {publisher} {John Wiley \& Sons},\
  \bibinfo {year} {2005})\BibitemShut {NoStop}%
\end{thebibliography}
\end{document}